\documentclass[sigconf,nonacm]{acmart}

\usepackage{booktabs} % To thicken table lines

\usepackage[english]{babel}
\usepackage{moresize}
\usepackage{amsmath}
\usepackage{algorithmic}
\usepackage{balance}
\usepackage{comment}
\usepackage{paralist}
\usepackage{bm}
\usepackage{svg}
\usepackage{pgfplots}
\usetikzlibrary{pgfplots.dateplot}

\usepackage{flushend}
\usepackage[english]{babel}
\usepackage{mathrsfs}
\usepackage{graphicx}

\usepackage{amssymb}
\usepackage{amsmath}
\usepackage{threeparttable} 
\usepackage{amsfonts}
\usepackage{url}
\usepackage{longtable}
\usepackage{rotating}
\usepackage{multirow}
\usepackage{mathrsfs}
\usepackage{subfigure}
\usepackage{subcaption} % ?????????
\usepackage{booktabs}
\usepackage{enumitem}
\usepackage[linesnumbered,algoruled,boxed,lined]{algorithm2e}
\usepackage{adjustbox}
\usepackage{hyperref}
\usepackage{pgfplots}
\usepackage{xcolor}
\usepackage{xspace}
\usepackage{color}
\usepackage{bm}
\usepackage{pifont}

\usetikzlibrary{pgfplots.dateplot}
\usepackage{filecontents}
\definecolor{tblue}{RGB}{31,119,180}
\definecolor{torange}{RGB}{255,127,14}
\definecolor{tgreen}{RGB}{44,160,44}
\definecolor{tred}{RGB}{214,39,40}
\definecolor{tpurple}{RGB}{148,103,189}

\newcommand{\hide}[1]{} %hide

\newcommand{\Slot}{\textsc{Slot}\xspace}

\newcommand{\revision}[1] {{#1}}

\newcommand{\symA}{$^{*}$}    % 星号
\newcommand{\symB}{$^{\dagger}$}  % 匕首符号

\copyrightyear{2025}
\acmYear{2025}
\setcopyright{acmlicensed}
\acmConference[CCS '25] {Proceedings of the 2025 ACM SIGSAC Conference on Computer and Communications Security}{ October 13--17, 2025}{Taipei, Taiwan.}
\acmBooktitle{Proceedings of the 2025 ACM SIGSAC Conference on Computer and Communications Security (CCS '25), October 13--17, 2025, Taipei, Taiwan}
\acmISBN{979-8-4007-1525-9/2025/10}
\acmDOI{10.1145/3719027.3744788}
\title{\Slot: Provenance-Driven APT Detection through Graph Reinforcement Learning}

\author{Wei Qiao\symA}
\affiliation{%
		\institution{Xidian University}
		\institution{State Key Laboratory of Integrated Services Networks (ISN)}
		\city{Xi'an}
		\country{China}
	}
\email{qiaoweiXidian@gmail.com}

\author{Yebo Feng\symA}
\affiliation{%
		\institution{Nanyang Technological University}
		\country{Singapore}
	}
\email{yebo.feng@ntu.edu.sg}

\author{Teng Li\symB}
\affiliation{%
		\institution{Xidian University}
		\institution{State Key Laboratory of Integrated Services Networks (ISN)}
		\city{Xi'an}
		\country{China}
	}
\email{litengxidian@gmail.com}

\author{Zhuo Ma\symB}
\affiliation{
		\institution{Xidian University}
		\city{Xi'an}
		\country{China}
	}
\email{mazhuo@mail.xidian.edu.cn}

\author{Yulong Shen}
\affiliation{%
		\institution{Xidian University}
		\city{Xi'an}
		\country{China}
	}
\email{ylshen@mail.xidian.edu.cn}

\author{Jianfeng Ma}
\affiliation{%
		\institution{Xidian University}
		\city{Xi'an}
		\country{China}
	}
\email{jfma@mail.xidian.edu.cn}

\author{Yang Liu}
\affiliation{%
		\institution{Nanyang Technological University}
		\country{Singapore}
	}
\email{yangliu@ntu.edu.sg}
\thanks{\symA Wei Qiao and Yebo Feng contributed equally to this research.}
\thanks{\symB Corresponding authors: Teng Li and Zhuo Ma}

\renewcommand{\shortauthors}{Wei Qiao, Yebo Feng, Teng Li, Zhuo Ma, Yulong Shen, Jianfeng Ma, Yang Liu}



\begin{document}

\begin{abstract}

Advanced Persistent Threats (APTs) represent sophisticated cyberattacks characterized by their ability to remain undetected within the victim system for extended periods, aiming to exfiltrate sensitive data or disrupt operations. Existing detection approaches often struggle to effectively identify these complex threats, construct the attack chain for defense facilitation, or resist adversarial attacks. To overcome these challenges, we propose \Slot, an advanced APT detection approach based on provenance graphs and graph reinforcement learning. \Slot excels in uncovering multi-level hidden relationships, such as causal, contextual, and indirect connections, among system behaviors through provenance graph mining. \revision{\Slot implements semi-supervised learning with limited labels through efficient label similarity computation, significantly enhancing both detection performance and model robustness. By pioneering the integration of graph reinforcement learning, \Slot dynamically adapts to new user activities and evolving attack strategies, enhancing its resilience against adversarial attacks. Additionally, \Slot automatically constructs the attack chain according to detected attacks with clustering algorithms, providing precise identification of attack paths and facilitating the development of defense strategies.} Evaluations with real-world datasets demonstrate \Slot's outstanding accuracy, efficiency, adaptability, and robustness in APT detection, with most metrics surpassing state-of-the-art methods. Additionally, case studies conducted to assess \Slot's effectiveness in supporting APT defense further establish it as a practical and reliable tool for cybersecurity protection.

\end{abstract}

\maketitle
\newcommand{\upp}[1]{\color[rgb]{0.117, 0.447, 0.999}#1}
\newcommand{\dow}[1]{\color[rgb]{0.753,0,0}#1}
\newcommand{\inc}[1]{\upp\mathbf{\blacktriangle #1\%}}
\newcommand{\dec}[1]{\dow\mathbf{\blacktriangledown #1\%}}

\section{Introduction}
\label{sec:intro}

An Advanced Persistent Threat (APT) is a sophisticated cyberattack in which a malicious actor gains unauthorized access to a network, remaining undetected for a long time to steal sensitive data or disrupt system operations~\cite{sharma2023advanced}. These attacks are highly targeted, using advanced techniques to maintain stealth and persistence within the victim's network~\cite{alshamrani2019survey}, resulting in severe damage to the victim's system. For example, the latest SektorCERT report from Denmark~\cite{sektorcert2023attack} reveals that Sandworm, an APT hacking group, launched coordinated attacks on 22 Danish energy companies, causing widespread disconnection of remote devices. Additionally, Symantec researchers have uncovered a four-month APT operation by Grayling~\cite{symantec2023grayling}, targeting manufacturing, IT, and biomedical sectors in Taiwan, the US, and Vietnam, involving theft of sensitive data.

To effectively combat APTs, it is crucial to detect them early with both timeliness and precision.
Traditional detection methods may struggle with these complex and prolonged attacks, often resulting in either inadequate accuracy or excessive time to produce detection results~\cite{feng2025unmasking,feng2023explainable}. Recently, researchers have increasingly utilized provenance graphs to trace APT activities, progressively achieving practical efficacy and efficiency.

Based on their methodologies, existing provenance-graph-based approaches for APT detection can be classified into three categories: \textbf{1) Statistics-based approaches}~\cite{liu2018towards, hassan2019nodoze, wang2020you} consider infrequent events as suspicious and use provenance graphs to model and identify these events.
However, these approaches focus solely on direct event connections, overlooking the deep semantics and hidden relationships within provenance graphs, which can result in a high rate of false positives and diminished reliability.
\textbf{2) Specification-based approaches}~\cite{hassan2020tactical, milajerdi2019holmes, hossain2017sleuth, milajerdi2019poirot, hossain2020combating} leverage expert knowledge to develop heuristic rules for APT detection. While this method effectively maintains a low false positive rate in cases it covers, creating these heuristic rules demands substantial prior expert knowledge, which must be distilled and refined by experts, making it a resource-intensive process.
\textbf{3) Learning-based approaches}~\cite{han2020unicorn, alsaheel2021atlas, pei2016hercule, liu2019log2vec, kapoor2021prov, wang2022threatrace, zengy2022shadewatcher, rehman2024flash, yang2023prographer} use various deep learning techniques to model APT attack patterns and system behaviors, subsequently utilizing these models for APT detection. These methods can achieve satisfactory detection performance, particularly with the integration of provenance graphs and graph neural network (GNN), which can uncover subtle connections between system events to enhance APT detection. However, they remain vulnerable to adversarial attacks, where attackers can easily mimic legitimate users to mislead the model into producing incorrect detection results.

To address the aforementioned gaps, we introduce \Slot, an advanced APT detection approach based on provenance graphs and graph reinforcement learning. \revision{Unlike existing methods, \Slot achieves enhanced APT detection accuracy by leveraging Latent Behavior Mining and Graph Reinforcement Learning to deeply excavate multi-level hidden relationships (e.g., causal dependencies, contextual interactions, and indirect attack traces) from system behavior data.} Additionally, by incorporating graph reinforcement learning, \Slot can autonomously learn and adapt to new user activities and attack strategies without relying on threat reports derived from expert insights, thereby enhancing its adaptability and resilience against adversarial attacks. Furthermore, \Slot can automatically construct the attack chain, accurately identifying the attack path, which simplifies the defense process.

\Slot is comprised of five integrated modules that work together to detect highly stealthy APT attacks and assist administrators in formulating effective defense strategies. (1) It begins with the Graph Construction Module, which processes system logs to create an initial provenance graph that reflects system behaviors. (2) The Latent Behavior Mining Module then applies attention mechanisms and graph transformation techniques to discover hidden relationships within the graph, enhancing the depth and comprehensiveness of subsequent analysis. (3) The Embedding Module utilizes a graph reinforcement learning algorithm (i.e., Bernoulli multi-armed bandit~\cite{zhang2022hierarchical}) to embed semantic and topological features of graph nodes as vectors, aggregates similar nodes, and produces updated feature vectors to filter out camouflaged entities, thereby improving accuracy and robustness. (4) The Threat Detection Model, which combines a multi-layer perceptron (MLP)~\cite{popescu2009multilayer} and Isolation Forest (iForest)~\cite{cheng2019outlier} to comprehensively identify APTs. (5) Finally, the Attack Chain Reconstruction module traces the attack pathway by clustering correlative nodes, aiding in the development of effective defense strategies.

We evaluated \Slot using three publicly available APT datasets from well-respected institutions and research communities. The evaluation results show that \Slot achieves an overall detection accuracy of approximately 99\%, outperforming all existing state-of-the-art (SOTA) techniques. Furthermore, \Slot completes the detection process more quickly. Additionally, we tested \Slot against adversarial strategies proposed by Goyal et al.~\cite{goyal2023sometimes}, which involve altering the neighborhood of nodes in the attack graph to resemble benign nodes. Leveraging its graph reinforcement learning mechanism, \Slot exhibits enhanced resistance compared to SOTA approaches. Ultimately, \Slot has proven its effectiveness in tracing attack pathways and aiding in the development of defense strategies in case studies.

\revision{
In summary, this paper makes the following contributions.
\begin{itemize}[leftmargin=*]
    \item We propose \Slot, an accurate method for Advanced Persistent Threat (APT) detection that maintains robustness in adversarial environments. It effectively identifies malicious activities within large-scale audit logs while reconstructing attack chains to support the development of targeted defense strategies.
    \item We propose an advanced graph mining technique capable of efficiently uncovering multi-level hidden relationships---such as causal, contextual, and indirect connections---within the graph.
    \item \Slot implements semi-supervised learning with limited labels through efficient label similarity computation, significantly enhancing both detection performance and model robustness.
    \item We integrate reinforcement learning into provenance graph analysis. By embedding semantic and topological features and leveraging the adaptive dynamics of reinforcement learning, \Slot effectively endures highly adversarial environments.
    \item We performed comprehensive evaluations using real-world datasets, and the results underscore \Slot's effectiveness in detecting APTs, its resilience against adversarial attacks, and its capability to support the development of effective defense strategies.
\end{itemize}
}
\section{Related Work and Its Limitations}
\begin{figure*}[ht]
	\centering
	\includegraphics[width=\textwidth]{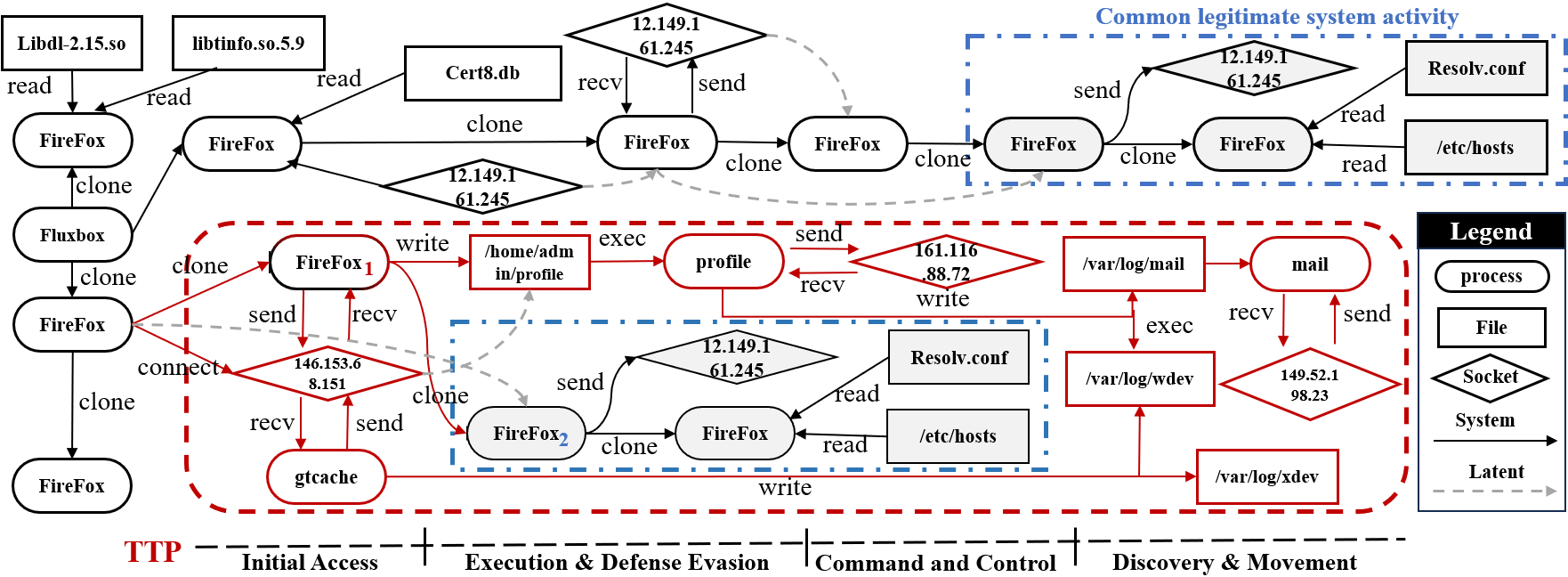}
\caption{An example of an attack provenance graph from the DARPA E3 dataset~\cite{DARPA}. Attackers launched a series of attacks exploiting the Firefox backdoor vulnerability, along with camouflage activities to cover up the attacks. \Slot effectively identifies all critical attack components, as indicated by the red nodes in the figure.}
	\label{fig:Attack Scenario}
\end{figure*}
\label{sec:relate}
This section categorizes and summarizes the related work on APT detection and graph reinforcement learning, and discusses their respective limitations.

\begin{table}[tph]
         \caption{Obstacles that limit the performance of APT detection approaches.}
		\tabcolsep=2.5pt
		\footnotesize
		\setlength{\belowcaptionskip}{1pt} % 
\resizebox{0.48\textwidth}{!}{
\begin{tabular}{lccccc}
\hline
\multicolumn{1}{c}{\textbf{Model}}       & \textbf{\begin{tabular}[c]{@{}c@{}}Threat \\ Report\end{tabular}} & \textbf{\begin{tabular}[c]{@{}c@{}}Semantic\\ Encoding\end{tabular}} & \textbf{\begin{tabular}[c]{@{}c@{}}Attack\\ Reconstruction\end{tabular}} & \textbf{\begin{tabular}[c]{@{}c@{}} \revision{Adversarial} \\ \revision{Robustness}\end{tabular}} & \textbf{Granularity} \\ \hline
\textbf{\Slot}                           & \textcolor{tred}{\ding{55}}                                        & \textcolor{tgreen}{\ding{51}}                                         & \textcolor{tgreen}{\ding{51}}                                             & \textcolor{tgreen}{\ding{51}}                                            & Node                 \\
Nodoze\cite{hassan2019nodoze}            & \textcolor{tred}{\ding{55}}                                        & \textcolor{tred}{\ding{55}}                                           & \textcolor{tred}{\ding{55}}                                               & \textcolor{tred}{\ding{55}}                                              & Node                 \\
Poirot\cite{milajerdi2019poirot}         & \textcolor{tgreen}{\ding{51}}                                      & \textcolor{tgreen}{\ding{51}}                                         & \textcolor{tgreen}{\ding{51}}                                             & \textcolor{tred}{\ding{55}}                                              & Graph                \\
Unicorn\cite{han2020unicorn}             & \textcolor{tred}{\ding{55}}                                        & \textcolor{tgreen}{\ding{51}}                                         & \textcolor{tred}{\ding{55}}                                               & \textcolor{tred}{\ding{55}}                                              & Graph                \\
ShadeWatcher\cite{zengy2022shadewatcher} & \textcolor{tred}{\ding{55}}                                        & \textcolor{tgreen}{\ding{51}}                                         & \textcolor{tred}{\ding{55}}                                               & \textcolor{tred}{\ding{55}}                                              & Edge                 \\
FLASH\cite{rehman2024flash}              & \textcolor{tred}{\ding{55}}                                        & \textcolor{tgreen}{\ding{51}}                                         & \textcolor{tgreen}{\ding{51}}                                             & \textcolor{tgreen}{\ding{51}}                                              & Node                 \\
MAGIC\cite{jia2024magic}                 & \textcolor{tred}{\ding{55}}                                        & \textcolor{tred}{\ding{55}}                                           & \textcolor{tred}{\ding{55}}                                               & \textcolor{tred}{\ding{55}}                                            & Node                 \\
ThreaTrace\cite{wang2022threatrace}      & \textcolor{tred}{\ding{55}}                                        & \textcolor{tred}{\ding{55}}                                           & \textcolor{tred}{\ding{55}}                                               & \textcolor{tred}{\ding{55}}                                            & Node                 \\ \hline
\end{tabular}
}
    \label{tab:Obstacles}
    
\end{table}

\subsection{APT Detection}
To effectively detect sophisticated APTs, provenance graphs are used to model system events and trace attacks. Recent detection methods can be classified into three categories~\cite{zengy2022shadewatcher}: statistics-based, specification-based, and learning-based methods. And it lists the existing limitations of current approaches in Table~\ref{tab:Obstacles}.

\textbf{Statistics-based approaches}: Statistical methods~\cite{hassan2019nodoze,liu2018towards,wang2020you} assume that attacks correlate with unusual system activities, quantifying suspiciousness based on interaction frequencies among system entities. However, rare events aren't always anomalies. For instance, in our scenario~\ref{fig:Attack Scenario}, Firefox loads the benign module $Org.chromium.iyhyah$ for the first time, which a statistical method flags as a false positive. This example shows that such methods fail to capture deep semantics and latent relationships, leading to high false positive rates.

\textbf{Specification-based approaches}: Specification-based methods match audit logs with threat report~\cite{milajerdi2019poirot, hassan2020tactical} or use expert knowledge for abnormality scoring~\cite{milajerdi2019holmes, hossain2020combating}, raising alerts when anomalies exceed a set threshold. While expert-driven detection generally achieves a low false positive rate, it requires specialized expertise to design effective strategies. As systems evolve and attack techniques grow more sophisticated, continuous expert intervention is needed, making it difficult for the system to adapt dynamically to changes in the network environment.

\textbf{Learning-based approaches}: Learning-based methods have proven effective for classification and anomaly detection by modeling system behaviors from logs. For example, Unicorn~\cite{han2020unicorn} uses graph similarity matching to identify anomalous graphs, but a single alert can involve thousands of logs, complicating verification. At the edge level, ShadeWatcher~\cite{zengy2022shadewatcher} models system interactions using GNNs, but high computational costs arise from numerous edges. At the node level, FLASH~\cite{rehman2024flash} and ThreaTrace~\cite{wang2022threatrace} utilize GNN to detect anomalies at the node level by learning the structural information of nodes and detecting deviations from this learned behavior. MAGIC~\cite{jia2024magic} utilizes masked autoencoder and KNN to identify abnormal nodes. While node-level detection effectively identifies anomalies, current methods fail to utilize heterogeneous entity information and overlook the impact of benign masquerading by neighboring nodes on embeddings. As exemplified in our attack scenario (Figure~\ref{fig:Attack Scenario}), embedding benign DNS resolution behaviors within the malicious red attack chain can disrupt the embedding representation of the root malicious node, "Firefox1", thereby enabling adversarial mimicry and evasion.

\subsection{Graph Reinforcement Learning}
Reinforcement Learning (RL) has recently achieved success in addressing challenges across various fields~\cite{feng2020application}, including robotics~\cite{levine2018learning}, gaming~\cite{chen2022game}, and natural language processing (NLP)~\cite{uc2023survey}. Researchers have found that RL methods enable effective exploration of the topological structures and attribute information of graphs by analyzing key components such as nodes, links, and subgraphs. SUGAR~\cite{sun2021sugar} employs Q-learning to adaptively select significant subgraphs to represent discriminative information of the graphs. Bachu et al.~\cite{bacciu2022explaining} develop perturbation strategies for local explanations of graph data by optimizing multi-objective scores. Other related works~\cite{yuan2021explainability, jiang2018graph} have demonstrated impressive performances in graph representation learning through RL, proving its efficacy in this domain. But currently, graph reinforcement learning has not yet been attempted in the field of log provenance detection.

\section{Problem Formalization}
\label{sec:Motivation}
% In this section, we introduce real attack scenarios from the DARPA~\cite{DARPA} dataset to highlight the limitations of existing provenance-based APT detection systems. Additionally, we define the threat scenarios that the \Slot must address and the objectives it aims to achieve. 
In this section, we present real-world attack scenarios from the DARPA dataset~\cite{DARPA} to highlight the limitations of current provenance-based APT detection systems. These scenarios demonstrate how existing systems may fail to detect complex and evolving attacks. We also define the specific threat scenarios that the \Slot system aims to address, focusing on improving detection accuracy, reducing false positives, and enhancing response times.
% \vspace{-3.75mm}
\subsection{Attack Scenario}
\label{sec: Attack}
Figure~\ref{fig:Attack Scenario} illustrates a Firefox backdoor attack. The red subgraph represents a simplified version of the original attack. A victim computer running the vulnerable Firefox version 54.0.1 unknowingly interacts with a malicious ad server located at 146.153.68.151. This server exploits a backdoor in Firefox, injecting the binary executable '$Drakon$'
1 into the process memory. '$Drakon$' subsequently spawns a new process with root privileges (/home/admin/profile), which connects to an attacker's server at 161.116.88.72, thus granting the attacker full access to the victim's computer.
%现有的基于学习APT检测通过学习溯源图中的行为模式来检测攻击（红色实体）和良性（黑色实体），并且系统中常见的DNS解析行为（蓝色子图）在多次学习中会被表达为更加xx的良性结构。但是APT攻击中，攻击者在长时间的信息渗透过程中，可能会掌握系统的常见良性活动，从而发动对抗性攻击，在攻击中嵌入图中蓝色框中的良性子结构，但并不影响其实际攻击逻辑。通过恶性进程多次执行常见良性活动，从而在检测过程中，诱导攻击实体嵌入表达为良性结果，实现攻击逃逸

The existing APT detection systems based on learning distinguish between attack activities (red entities) and benign behaviors (black entities) by analyzing behavior patterns in provenance graphs. Common DNS resolution behaviors (blue subgraph) can be classified more clearly as benign structures after repeated learning sessions. However, in APT attacks, attackers may familiarize themselves with and mimic common benign activities of the system over a long period of information infiltration, thereby launching adversarial attacks that embed these benign substructures into their attack tactics without altering the actual attack logic. By repeatedly performing these common benign activities, malicious processes can be misjudged as benign during the detection process, thus achieving attack evasion.

\subsection{Threat Model}
\label{sec:Threat Model}
Building on prior APT detection research \cite{wang2022threatrace, rehman2024flash, han2020unicorn, zengy2022shadewatcher}, our study focuses on scenarios where attackers exploit software vulnerabilities and communication backdoors for system control and persistence. We exclude hardware trojans and side-channel attacks, which are undetectable through system audits. We assume attackers know the target hosts' benign activities, enabling mimicry attacks. Additionally, we consider audit log data secure, supported by robust security tracing~\cite{pasquier2017practical} and anti-tampering~\cite{paccagnella2020custos, ahmad2022hardlog}, making our provenance graphs reliable for effective threat detection and analysis.

\subsection{Design Goals}
\label{sec:Design Goals}
To defend against APT attacks, \Slot models and analyzes system call data recorded by audit logs. The ultimate goal of \Slot is to provide security analysts with more effective and streamlined attack insights, thereby accelerating the alert handling process. We believe that APT detection should achieve the following objectives when raising alarms: (1) For the complex multi-stage characteristics of APT attacks, \Slot must have precise detection capabilities to identify attack-related entities; (2) Amid vast volumes of logs, \Slot should maintain a low false positive rate to prevent alert fatigue; (3) In the face of highly covert adversarial attacks, \Slot must be highly robust, capable of dealing with attackers' camouflaging behaviors; (4) Finally, \Slot should provide security analysts with more streamlined and effective traceable analysis results, enabling the validation of alarm events within a reasonable time cost.

\section{Methodology}
\label{sec:solution}

\subsection{Overview}
\begin{figure*}[htp]
	\centering
	\includegraphics[width=\textwidth]{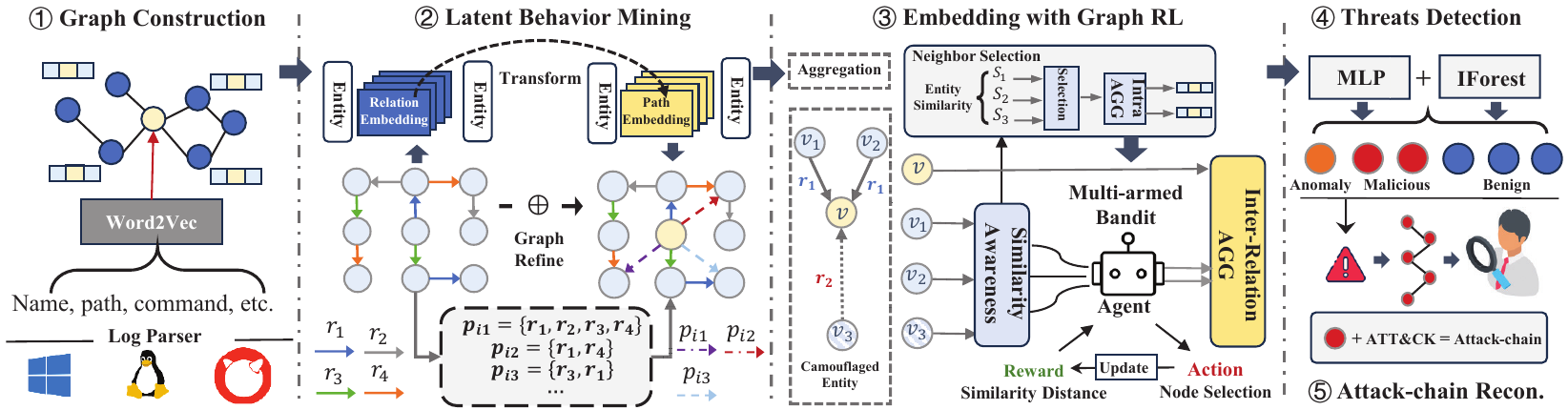}
	
	\caption{\revision{The workflow of \Slot. \Slot initializes event node features through semantic embedding, enhances graph relationships via latent relation mining, and then employs graph reinforcement learning to achieve robust node embedding representations, ultimately accomplishing malicious activity detection and attack reconstruction.}}
	\label{fig:framework}
	
\end{figure*}
\Slot is a fine-grained, adversarial threat provenance scheme. It leverages attention mechanisms and graph transformation techniques to uncover deep hidden relationships, and then combines reinforcement learning with the provenance graph to guide relational aggregation in GNN. \Slot features flexible policy extension capabilities and dynamic adaptability, allowing it to address adversarial attacks in real-world scenarios. \Slot not only provides malicious detection through model learning but also proposes a method for considering unknown anomalies. Finally,  By integrating LPA and ATT\&CK~\cite{mitre_attack} encoding, it offers a concise and low false-positive attack chain, effectively accelerating manual verification. Figure~\ref{fig:framework} depicts \Slot' architecture consisting of five major components:

\noindent\textbf{\textcircled{1} Graph Construction (\S~\ref{Graph Construction}).} \Slot constructs a graph based on entity relationships from system call logs and extracts semantic information from the logs as initial features for the nodes.

\noindent\textbf{\textcircled{2} Latent Behavior Mining (\S~\ref{Behavior Mining}).} \Slot further explored the behavioral relationships on the original system provenance graph. It combined system call relationships to construct deeper path connections. This approach enriches event behaviors and accelerates the propagation of information within the graph.

\noindent\textbf{\textcircled{3} Embedding with Graph Reinforcement Learning (\S~\ref{Reinforcement Learning}).} 
% 在图嵌入阶段，\Slot 通过图表示学习来刻画系统的行为模式。\Slot 通过考虑节点语义特征和局部拓扑结构来计算溯源图上实体间的相似性，然后利用强化学习通过相似性来指导GNN关系的选择和聚合，从而实现对系统行为的建模。
In the graph embedding stage, \Slot captures system behavior patterns through graph reinforcement learning. Specifically, \Slot combines the semantic features and the topological
features of nodes to calculate the similarity between entities in the provenance graph. Subsequently, it uses a reinforcement learning strategy to guide the selection and aggregation of relationships in the GNN based on this similarity, thereby achieving effective modeling of system behavior and obtaining the true feature vectors of the nodes.

\noindent\textbf{\textcircled{4} Threats Detection (\S~\ref{Threats Detection}).} Using the trained model, \Slot performs node representation on the test logs. For previously learned behavior patterns, \Slot can directly classify node vectors as benign or malicious. However, to address unknown behaviors in APT attacks, \Slot uses anomaly detection to identify entities that deviate from known behavior patterns. \Slot treats both malicious entities and anomalous entities as attack activities.

\noindent\textbf{\textcircled{5} Attack-chain Reconstruction (\S~\ref{Attack-chain Reconstruction}).} To assist security analysts in alert verification, \Slot constructs scattered nodes into a complete attack activity chain, eliminating the need for tracing across thousands of nodes. \Slot classifies attack nodes into different attack chains by clustering, using the TTP (Tactics, Techniques, and Procedures) phase tags from the ATT\&CK framework~\cite{mitre_attack} along with the feature embeddings of the nodes.

% \Slot is a fine-grained, adversarial threat provenance scheme. It utilizes a deep learning method based on reinforcement learning, focused on deeper latent relationship behaviors for attack detection. Moreover, \Slot features flexible policy extension capabilities and dynamic adaptability to real-world adversarial attacks. Figure~\ref{fig:framework} provides an architectural overview of \Slot, which consists of four main components: 1) Graph Construction \& Latent Relationship Mining, 2) Provenance Graph Representation Learning, 3) Threats Detection, and 4) Attack-chain Reconstruction.

\subsection{Graph Construction}
\label{Graph Construction}
\begin{table}[hpt]
\caption{System behaviors extracted from audit logs.}
\label{tab:System Behavior}
\resizebox{0.48\textwidth}{!}{
\begin{tabular}{ll}
\hline
\multicolumn{1}{c}{System Behavior}            & \multicolumn{1}{c}{Relation Description}                \\ \hline
Process $\rightarrow$ R1 $\rightarrow$ Process & "R1": "fork", "execute", "exit", "clone", etc. \\
Process $\rightarrow$ R2 $\rightarrow$ File    & "R2": "read", "open", "close", "write", etc.   \\
Process $\rightarrow$ R3 $\rightarrow$ Netflow & "R3": "connect", "send", "recv", "write", etc. \\
Process $\rightarrow$ R4 $\rightarrow$ Memory  & "R4": "read", "mprotect", "mmap", etc.         \\ \hline
\end{tabular}
}

\label{System Behavior.}
\end{table}
\Slot constructs a comprehensive system provenance graph using audit data collected from logging infrastructures such as Windows ETW~\cite{microsoft_event_tracing}, Linux Audit~\cite{linux_auditd}, and CamFlow~\cite{pasquier2018runtime}. It adheres to the existing definitions of provenance graphs and employs recent graph preprocessing techniques~\cite{wang2022threatrace, rehman2024flash}, abstracting events into subjects, objects, and relationships, as illustrated in Table~\ref{tab:System Behavior}. Here, source nodes like Process/Thread represent subjects, and target nodes like Files or sockets represent objects, with system call events serving as the relationships to form graph $\mathcal{G} (S, R, O)$. In addition to focusing on system behaviors, \Slot further enriches the relationships between behaviors and explores multi-hop latent relations in Section~\ref{Behavior Mining}. \revision{Regarding node features, \Slot fully considers meaningful semantic information such as process names and command-line parameters for process nodes, file paths for file nodes, network IP addresses and ports for netflow nodes. \Slot effectively mines deep semantic patterns in provenance logs by constructing node "summary sentences" through aggregating semantic attributes (process names, file paths, IPs) and causal events (system calls) from 1-hop neighbors, incorporating timestamp-sorted event sequences with positional encoding to preserve temporal relationships, and leveraging the Word2Vec model~\cite{mikolov2013efficient} to learn vector representations of semantic features.}

\subsection{\bf Latent Behavior Mining}
\label{Behavior Mining}
\begin{figure}[htp!]
    \centering
    \includegraphics[width=0.45\textwidth,height=3.5cm]{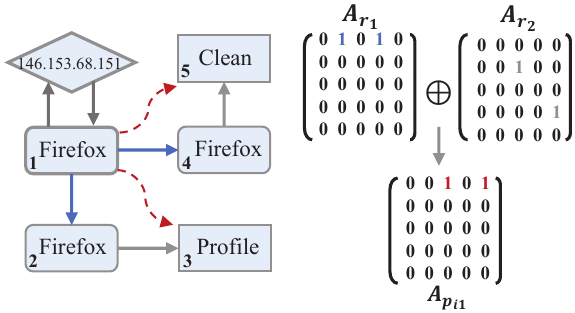}

    \caption{Latent behavior adjacency matrix construction, with colored edges indicating different relationships.}
    \label{fig:Latent}

\end{figure}
\revision{A complete user behavior or attack campaign typically involves multiple system calls, which manifest as multi-hop node relationships in provenance graphs. To accelerate node information propagation and uncover implicit dependencies between distant nodes, \Slot introduces a Latent Relation Mining Module. This module automatically constructs hidden relationships across multi-hop nodes through a relational attention mechanism, prioritizing causal chains and attack-relevant contextual patterns.}

In the mining of latent relationships, \Slot leverages attention mechanisms~\cite{velivckovic2017graph} and graph transformation~\cite{xie2022efficiency} techniques to automatically mine advanced behavioral paths. For basic system call behavior $r_i$, it mines deep event-behavior relationships $p_{ik} \in P_i$, and then integrates low-dimensional overall system relationships with path-based advanced behavioral relationships to generate a new provenance graph. We define $p_{ik} = {r_{i1}, r_{i2},...,r_{iL}}$, represented as an $L$-hop event path, where each $r_{il} \in p_{ik}$ is a basic system call behavior in the event path. In the path generation module, we use multiple attention mechanisms to softly select $L$ relationships and combine these relationships to form path $P_{ik}$. Specifically, for sub-behaviors within events, we generate corresponding relationship embedding using attention mechanisms: $\mathbf{r}_{il}=\sum_{j=1}^n\alpha_{ij}^l\mathbf{r}_j$, where $\alpha_{ij}^l$ is calculated through scores normalized by the softmax function, which is computed as follows: 
\begin{align}
\alpha_{ij}^l=\frac{\exp(\mathbf{r}_j^T\sigma(\mathbf{r}_i\mathbf{W}^l+\mathbf{b}^l))}{\sum_{j'=1}^n\exp(\mathbf{r}_{j'}^T\sigma(\mathbf{r}_i\mathbf{W}^l+\mathbf{b}^l))},
\end{align}
where $W$ and $b$ are parameters used in the path generation process, and $\sigma$ represents the activation function. By incorporating attention mechanisms, the model can effectively select multi-hop paths most relevant to a given relationship $r_i$, thereby enhancing the representation of complex relationships between entities in the provenance graph.
For each path $p_{ik}$ composed of $L$ hops, its adjacency matrix $A_{p_{ki}}$ is obtained by the product of the adjacency matrices of the corresponding relationships: $A_{p_{ik}}=A_{r_{i1}}\cdot A_{r_{i2}}\ldots A_{r_{iL}}$, where each $A_{r_{il}}$ is the adjacency matrix for the relationship $r_{il}$. Direct computation of this product can be resource-intensive, so our method uses an efficient representation to approximate this process. First, each relationship's adjacency matrix is approximated by $A_{ril}\approx E\cdot\mathrm{diag}(r_{il})\cdot E^{T}$, where $E$ represents the matrix of entity embeddings, and $diag(r_{il})$ transforms the relationship embedding $r_{il}$ into a diagonal matrix form. As shown in Figure~\ref{fig:Latent}, the adjacency matrix for the path $A_{p_{ik}}$ can be calculated as follows to capture the composite effect of multiple relationships along the path:
\begin{equation}
    \begin{aligned}A_{p_{ik}} & \approx(\mathbf{E}\cdot\mathrm{diag}(\mathbf{r}_{i1})\cdot \mathbf{E}^{T})\cdot\ldots\cdot(\mathbf{E}\cdot\mathrm{diag}(\mathbf{r}_{iL})\cdot \mathbf{E}^{T})\\  & \approx\mathrm{diag}(\mathbf{r}_{i1})\cdot\mathbf{E}^{T}\cdots\mathbf{E}\cdot\mathrm{diag}(\mathbf{r}_{iL})\\  & \approx\mathbf{E}\cdot\mathbf{p}_{ik}\cdot\mathbf{E}^{T}.\end{aligned}
\end{equation}

\subsection{Embedding with Graph Reinforcement Learning}
\label{Reinforcement Learning}
\Slot utilizes graph reinforcement learning to obtain high-quality node embeddings from the provenance graph. The embedding module consists of three stages: 1) In the similarity-aware phase, the node's semantic features and topological features are embedded into vectors. 2) Based on the similarity calculation of node feature vectors, we design an adaptive Bandit neighbor selector using reinforcement learning~\cite{peng2021reinforced}. Finally, we aggregate features by integrating multiple relationships (including system call and latent relationships) to generate updated node feature vectors.

\subsubsection{\bf Feature-Topology Similarity Awareness}
\begin{figure}[htp]
	\centering
	\includegraphics[width=0.46\textwidth]{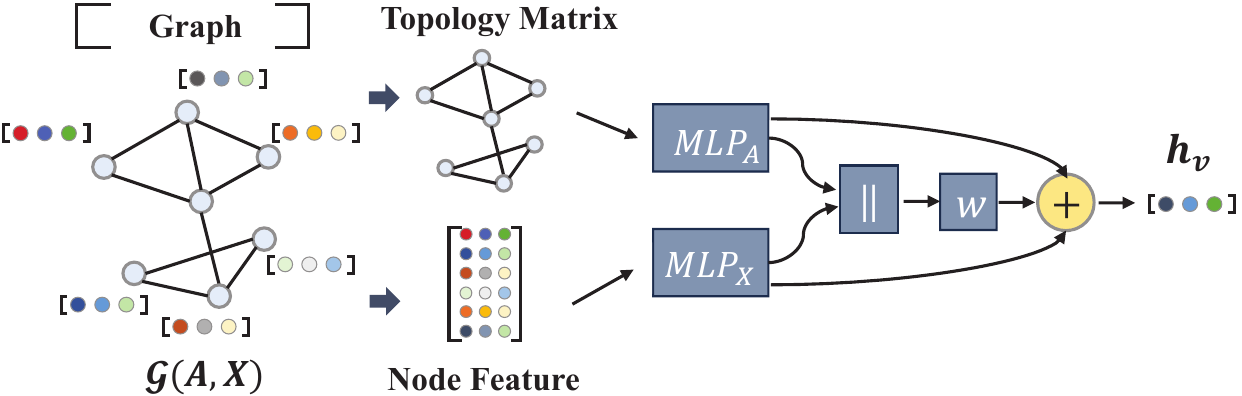}
	
	\caption{Embed semantic features and topological features as vectors.}
	\label{fig:Similarity}
	
\end{figure}
Previous research has explored various types of attacker camouflages from both behavioral~\cite{ofori2021topological, jiang2022camouflaged} and semantic~\cite{wen2020asa} perspectives. These camouflages can make the features of malicious activities similar to those of benign entities, further inducing GNN to generate incorrect node embeddings. To address these issues of node feature camouflage, we believe that an effective similarity metric is necessary to filter out disguised neighbors before applying GNNs. \Slot achieves lightweight node similarity awareness by simultaneously considering node attribute features and topological features~\cite{lim2021large} shown in Figure~\ref{fig:Similarity}.

\noindent \textbf{MLP on node features.} A simple approach to node classification is to ignore graph topology and simply train an MLP on node features. Research~\cite{zhu2020beyond} has demonstrated that MLPs can actually perform relatively well on heterogeneous graphs-achieving higher or roughly equivalent efficacy compared to various GNNs. Thus, we represent the node features as follows:
\begin{equation}
    \mathbf{h_X}=\mathrm{MLP_X}(\mathbf{X})\in\mathbb{R}^{d\times n},
\end{equation}
where $\mathbb{R}^{d\times n}$ denote the matrix of node features with input dimension d.

\noindent \textbf{LINK regression on graph topology.} Another extreme is LINK~\cite{zheleva2009join}, a simple baseline that solely utilizes graph topology. We extend LINK to compute feature embeddings for the topological relationship $A$ as follows:
\begin{equation}
    \mathbf{h_A}=\mathrm{MLP_A}(\mathbf{A})\in\mathbb{R}^{d\times n}.
\end{equation}
Finally, we let $[h_1; h_2]$ denote concatenation of vectors $h_1$ and $h_2$ and map the entity feature vectors as:
\begin{equation}
    \mathbf{h_v}=\mathrm{MLP}_f\left(\sigma\left(\mathbf{W}[\mathbf{h}_\mathbf{A};\mathbf{h}_\mathbf{X}]+\mathbf{h}_\mathbf{A}+\mathbf{h}_\mathbf{X}\right)\right).
\end{equation}

% 我们使用两个节点的预测结果之间的 l1 距离作为它们的相似性度量。
We use the L1 distance between the feature vectors of two nodes as their measure of similarity. For a center node v under relation r at the $l-th$ layer and edge $(v,v')\in\mathcal{E}_{r}^{(l-1)}$, we can define the similarity measure as:
\begin{align}
    \mathcal{D}^{(l)}(v,v^{\prime}) = \left\|\sigma\left(MLP^{(l)}(\mathbf{h}_v^{(l-1)})\right)-\sigma\left(MLP^{(l)}(\mathbf{h}_{v^{\prime}}^{(l-1)})\right)\right\|_1.
\end{align}

\subsubsection{\bf Adaptive Bandit Neighbor Selector}
Given relation camouflage, attackers may connect to varying numbers of benign entities under different relationships~\cite{goyal2023sometimes}. We should select similar neighbors (i.e., filter disguised neighbors) to enhance the capability of GNN. To adaptively select appropriate neighbors, we utilize a similarity-aware neighbor selector to filter out nodes exhibiting inappropriate behaviors due to adversarial actions or inaccurate feature extraction~\cite{peng2021reinforced}. More specifically, for each central node, the selector utilizes Top-p sampling and adaptive filtering thresholds to construct similar neighbors under each relationship.

\noindent \textbf{Top-p Sampling.} 

Before aggregating information from the central node $v$ and its neighbors, we perform Top-p sampling to filter dissimilar neighbors based on different relationships. The filtering threshold $p_r^l$ for the $l-th$ layer relationship $r$ is defined within the range $[0,1]$, representing the selection ratio from all neighbors. 

\noindent \textbf{Finding the Optimal Thresholds with RL.} 
\begin{figure}[htp]
	\centering
	\includegraphics[width=0.36\textwidth]{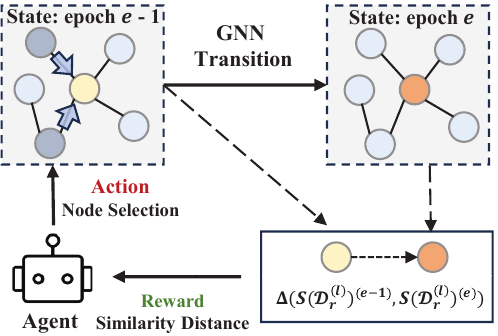}
	
	\caption {RL-guided graph neighbor node selection.}
	\label{fig:graphRF}
	
\end{figure}

The logic behind reinforced neighborhood selection is to mine appropriate neighbor nodes for the central node. Previous GNN-based Works~\cite{wang2022threatrace, rehman2024flash} are built on homogeneous benchmark graphs, devoid of noise caused by disguised attackers. Inspired by neighborhood sampling~\cite{peng2021reinforced}, we found that the scenario of resisting neighborhood interference satisfies the conditions for applying a Bernoulli multi-armed bandit reinforcement learning approach, allowing for the adaptive selection of suitable neighborhoods for each central node. We represent the RL module as a Markov Decision Process $MDP<A,S,R,T,F>$, used for the filtering threshold of relationships. $A$ is the action space, $S$ is the node state, $R$ is the reward function, $T$ is the state transition, and $F$ is the termination condition.
\begin{itemize}[leftmargin=*]
    \item {\textbf{Action}} The action in the RL model dictates the updates to $p_r^{(l)}$ according to the received rewards. Given that $p_r^{(l)}$ falls within the range $[0, 1]$, we define the action $a_r^{(l)}$ as modifying $p_r^{(l)}$ by adding or subtracting a fixed increment $\tau\in[0,1]$.
    \item {\textbf{State}} Due to the inability to use the GNN's loss as the environment state, we use the average node distance, computed through similarity-aware calculations, as the state. The average neighbor distance in train node $\mathcal{V}_{train}$ at layer $l$ under relation $r$ during epoch $e$ is:
    \begin{equation}
    \mathcal{S}(\mathcal{D}_r^{(l)})^{(e)}=\frac{\sum_{v\in\mathcal{V}_{train}}\mathcal{D}_r^{(l)}(v,v')^{(e)}}{|\mathcal{V}_{train}|}.
    \end{equation}
    \item {\textbf{Reward}} The optimal $p(l)^r$ at layer $l$ under relation $r$ aims to find the most similar neighbors to the central node. \revision{A decrease in the average neighbor distance signifies that the selected neighbors exhibit higher similarity to the central node.} We design a binary reward based on the difference in average distances between two consecutive epochs. We define the reward for epoch $e$ as:
    \begin{equation}
    	f(p_r^{(l)},a_r^{(l)})^{(e)}=\left\{\begin{array}{l}+1,\mathcal{S}(\mathcal{D}_r^{(l)})^{(e-1)}-\mathcal{S}(\mathcal{D}_r^{(l)})^{(e)}\geq0,\\-1,\mathcal{S}(\mathcal{D}_r^{(l)})^{(e-1)}-\mathcal{S}(\mathcal{D}_r^{(l)})^{(e)}<0.\end{array}\right.
    \end{equation}
	When the distance between two consecutive epochs is positive, the reward is positive; otherwise, the reward is negative. Finally, we greedily update the actions based on the rewards.
    \item {\textbf{Transition}} State transitions are achieved through the forward propagation of GNN. When the neighbor selection threshold changes, the embeddings of the nodes are updated, leading to an update of the entire system's state.
    \item {\textbf{Terminal}} We define the terminal condition for RL as:
    \begin{equation}
    \sum_{e-10}^{e}f(p_{r}^{(l)},a_{r}^{(l)})^{(e)}|\leq2, where \quad e\geq10.
    \label{terminal}
    \end{equation}
    It means that the RL converges in the recent ten epochs and indicates an optimal threshold $p_r^{(l)}$ is discovered.
\end{itemize}

\subsubsection{\bf Contextual Relationship Integrator}
After filtering the neighbors for each relationship through reinforcement learning, we directly use the optimal filtering threshold $p_r^{(l)}$, learned from the RL process as the aggregation weight between relationships. We then use a GNN to aggregate information from neighbors across different relationships, generating updated node feature vectors. Formally, at the $l-th$ layer for relationship $r$, after applying top-p sampling, the neighbor aggregation for node $v$ is defined as follows:
\begin{equation}
\mathbf{h}_{v,r}^{(l)}=\mathrm{ReLU}\left(\mathrm{GNN}_{r}^{(l)}\left(\left\{\mathbf{h}_{v^{\prime}}^{(l-1)}:(v,v^{\prime})\in\mathcal{E}_{r}^{(l)}\right\}\right)\right).
\end{equation}
Then, we define the inter-relation aggregation as follows:
\begin{equation}
\mathbf{h}_{v}^{(l)}=\mathrm{ReLU}\left(\mathrm{GNN}_{all}^{(l)}\left(\mathbf{h}_{v}^{(l-1)}\oplus\{p_{r}^{(l)}\cdot\mathbf{h}_{v,r}^{(l)}\}|_{r=1}^{R}\right)\right).
\end{equation}

\subsection{Threats Detection} 
\label{Threats Detection}
%在获得系统实体的表示之后，我们进行威胁检测学习，将系统实体分为正常和恶性。
After obtaining the embeddings of the system entities, we perform threat detection using a trained MLP and anomaly-based Isolation Forest, classifying the system entities into benign, malicious, and anomalous categories.

\noindent \textbf{Learning Detection.} Existing node-level approaches~\cite{rehman2024flash,wang2022threatrace,jia2024magic}, whether based on anomaly-based unsupervised detection or self-supervised methods relying on data structure or attributes, have shown very high false positive rates. Gui et al.~\cite{gui2024survey} argued that in highly adversarial and data-imbalanced scenarios, semi-supervised learning with a small amount of labeled data can more effectively capture imbalanced data patterns, thereby increasing the embedding distance between benign and malicious behavior patterns. To classify the embedded features, we trained an MLP classifier. Slot uses only 10\% of the attack data to assist in characterizing the benign patterns, and through oversampling, semi-supervised learning is performed with the benign training set. For jointly training the similarity measure with GNNs, a heuristic approach involves appending it as a new layer before the GNN aggregation layer. We define the cross-entropy loss for the MLP at the first layer as follows:
\begin{equation}
    \mathcal{L}_{\mathrm{Simi}}^{(1)}=\sum_{\upsilon\in\mathcal{V}}-\log\left(y_{\upsilon}\cdot\sigma\left(MLP^{(l)}(\mathbf{h}_{\upsilon}^{(l)})\right)\right).
\end{equation}
For each node v, its final embedding is the output of the GNN at the last layer $\textbf{z}_v = \textbf{h}^{(L)}_v$ . We can define the loss of GNN as a cross-entropy loss function:
\begin{equation}
    \mathcal{L}_{\mathrm{GNN}}=\sum_{v\in\mathcal{V}}-\log\left(y_{v}\cdot\sigma(MLP(\mathbf{z}_{v}))\right).
\end{equation}
By combining the losses of the similarity modeling and the GNN modeling, we minimize the following objective function to learn parameters in our threats detection model:
\begin{equation}
\mathcal{L}=\mathcal{L}_{\mathrm{GNN}}+\lambda_{1}\mathcal{L}_{\mathrm{Simi}}^{(1)}+\lambda_{2}||\Theta||_{2},
\end{equation}
where $||\Theta||_2$ is the $L2$-norm of all model parameters, $\lambda_{1}$ and $\lambda_{2}$ are weighting parameters. Finally, we can classify the node feature vectors using the trained MLP.

\noindent\textbf{Anomaly Detection.}
Learning-based scheme is effective in detecting known benign or malicious patterns but struggles with identifying unknown anomalies. To address this limitation, we have additionally introduced an unsupervised detector on top of the supervised learning classifier to identify unknown anomalies, or outliers. The key intuition here is that anomalous entities, compared to learned behavior patterns, tend to be sparse and distant from other data points.

\Slot utilizes an Isolation Forest~\cite{cheng2019outlier} to partition node feature vectors and compute node anomaly scores. Initially, the feature vector $h$ in the detection representation is normalized as $\mathbf{h}^{\prime}=\frac{\mathbf{h}-\mu_{\mathbf{h}}}{\sigma_{\mathbf{h}}}$. Subsequently, the node anomaly score is calculated as:
\begin{equation}Score(\mathbf{h},n)=2^{-\frac{E(l(\mathbf{h}^{\prime}))}{c(n)}},\end{equation}
where $n$ represents the total number of samples, and $E(l(\mathbf{h}^{\prime}))$ denotes the average path length of the data point across all trees, which indicates the average depth at which the data point is isolated (i.e., the path ends) in the tree. $c(n)$ represents the average path length of a data point in a completely random tree.

After parallel detection, \Slot combines the results of both classifiers to make the final decision. If the supervised learning model makes a high-confidence prediction about a behavior (either benign or malicious), that prediction is adopted. On the other hand, if the anomaly detection model indicates that the behavior might be anomalous, it is labeled as anomalous.

%基于学习的方案只能检测已知的良性或攻击模式，难以对未知异常进行检测，对此，我们额外增加了一个无监督检测器，来提供对于未知异常（离群点）的判别。一个简单的直觉是，异常于已学习的行为模式的实体通常是稀疏的且与其他数据点相隔较远，Slot通过孤立森林，对节点特征进行切分，计算节点异常分数。首先将检测表示中的特征向量h进行标准化处理$h^{\prime}=\frac{h-\mu_h}{\sigma_h}$，然后计算节点异常分数：\begin{equation}s(h,n)=2^{-\frac{E(l(h^{\prime}))}{c(n)}}\end{equation}, 其中n是样本总数，l(h^{\prime}) 是数据点在所有数中的平均路径长度，即数据点在树中被孤立（路径结束）的平均深度。$c(n)$ 是数据点在完全随机树中的平均路径长度。
%Slot在并行检测后，结合两种分类器的结果来进行最终决策。如果监督学习模型对一个行为有高置信度的预测（良性或恶性），则采用该预测。反之通过异常检测模型提示此行为可能是异常，那么将其标记为未知或异常。

\subsection{Attack-chain Reconstruction}
\label{Attack-chain Reconstruction}
%%% 考虑到基于节点的告警会分散在极大的溯源图中，验证一个事件行为需要对节点向前或向后溯源调查, 分散的节点于对于审核人员的分析是很不便的，且基于节点级的告警伴随大量的假阳节点，会带来严重的告警疲劳问题。
%为了解决这一挑战，我们提出了一种基于告警节点的重构方案，与先前采用直接采用聚类或路径关联的方式不同，\Slot首次在攻击链重构中引入了ATT&CK框架，映射ATT&CK框架中的攻击技术，为每个节点分配适当的TTP（Tactics, Techniques, and Procedures）编码。\Slot用独热编码表示TTP，如果ATT&CK框架中有N种不同的TTP，则每个TTP编码为N维向量，其中相应的TTP位置为1，其余为0，每种技术或策略对应一种编码。我们将节点特征与TTP编码连接获得节点初始标签xi=[hi;ti], 然后通过LPA算法，更新节点标签：
%\begin{equation}
%	\mathbf{y}_i^{(t+1)}=\text{mode}(\{\mathbf{y}_j^{(t)}:j\in\mathcal{N}(i)\})
%\end{equation}
%其中yi是节点第i次迭代的标签，N(i) 是节点的邻接节点集合，mode表示选择最频繁的标签。重复标签传播过程，直到所有节点的标签不再发生变化。通过TTP模式结合可以有效过滤攻击链中的假阳性节点，并生成最终的攻击路径。

%
%在重构中保证了社区的信息聚合度，过滤掉与社区攻击模式不相关的路径，并最终生成可能的攻击路径。
%
%
%从攻击初始节点（源节点）到已发现的异常节点（目标节点）之间的路径，来重构攻击链。它使用深度优先搜索（DFS）算法，同时结合了一些时间顺序（路径节点发生顺序），攻击模式匹配（将路径与Holmos中的攻击阶段比对，路径与已知模式越相似，得分越高），以过滤掉不相关的路径，并最终生成可能的攻击路径。算法1说明了Alert-chain的生成过程。
%
%
Considering that node-based alerts are dispersed in extremely large provenance graphs, verifying an event activity requires a forward or backward traceability investigation of nodes. Dispersed nodes are inconvenient for security analysts to validate the analysis, and node-level based alerts are accompanied by a large number of false-positive nodes, which can lead to severe alert fatigue.

To address this challenge, we propose a novel reconstruction scheme based on alert nodes, which, unlike previous approaches~\cite{rehman2024flash, cheng2023kairos} that directly employed clustering or path association, incorporates the ATT\&CK framework~\cite{mitre_attack} into attack chain reconstruction for the first time. This method involves mapping attack techniques from the ATT\&CK framework and assigning appropriate TTP (Tactics, Techniques, and Procedures) codes $\mathbf{t}_i$ to each node. We represent TTPs using one-hot encoding; if the ATT\&CK framework includes N distinct TTPs, each TTP is encoded as an N-dimensional vector, with the position corresponding to the TTP set to 1 and the rest set to 0, with each technique or tactic corresponding to one code. We concatenate the node features with the TTP codes to obtain the initial node label $x_i=[\mathbf{h}_i;\mathbf{t}_i]$, and then update the node labels using the Label Propagation Algorithm (LPA): 
\begin{equation} 
	\mathbf{y}_i^{(t+1)}=\text{mode}({\mathbf{y}_j^{(t)}\in\mathcal{N}(i)}). 
\end{equation} 
where $y_i$ represents the label of node $i$ during the $i-th$ iteration, $N(i)$ denotes the set of neighboring nodes of $i$, and mode indicates the selection of the most frequent label. The label propagation process is repeated until the labels of all nodes no longer change. This combination of TTP patterns effectively filters out false positives within the attack chain and generates the final attack path.

\section{Evaluation}
\label{sec:eval}
In this section, experiments are performed to validate \Slot's advantage and answer the following key research questions:
\begin{itemize}[leftmargin=*]
\item  \textbf{RQ1}: Can similarity awareness provided by a small number of labels enable effective APT detection under semi-supervision?
\item \textbf{RQ2}: How effective is \Slot in detecting APTs compared to current state-of-the-art appraoches?
\item \textbf{RQ3}: What is the system overhead associated with using \Slot?
\item \textbf{RQ4}: How effective are the components of our \Slot in achieving their intended functions?
\item \textbf{RQ5}: How do different hyperparameters influence the detection capabilities of \Slot?
\item \textbf{RQ6}: How resilient is \Slot to adversarial attacks?
\item \textbf{RQ7}: How effectively does \Slot facilitate the manual validation of alerts?

\end{itemize}

\subsection{Experimental Setups}
\subsubsection{\bf Datasets}
The evaluation of \Slot was carried out using three open-source datasets: StreamSpot~\cite{Thestreamspotdataset}, Unicorn Wget~\cite{han2020unicorn} and DARPA E3~\cite{DARPA}. Both StreamSpot and Unicorn consist of batch audit logs generated in controlled environments. The StreamSpot dataset includes 600 batch logs monitoring system calls across six distinct scenarios, five of which simulate benign user behavior, while one represents a drive-by download attack. The Unicorn dataset contains 150 batch logs collected via Camflow~\cite{pasquier2018runtime}, with 125 logs corresponding to benign activity and 25 involving supply chain attacks. These attacks are stealthy in nature, designed to mimic normal system workflows, and are challenging to detect. The DARPA dataset is part of the DARPA Transparent Computing program, gathered during adversarial engagements in enterprise networks. The Red Team conducted APT attacks using various vulnerabilities to exfiltrate sensitive information, while the Blue Team focused on identifying these attacks through network host audits and causal analysis. We conducted experimental comparisons using datasets commonly shared across multiple SOTA algorithms~\cite{wang2022threatrace,rehman2024flash,jia2024magic,zengy2022shadewatcher} and recognized open-source labels~\cite{rehman2024flash,jia2024magic}.

\begin{table}[htp]
	\caption{Overview of the experimental datasets.}
	\centering
	\tabcolsep=3.5pt
	% \small
	%	%\setlength{\abovecaptionskip}{-27pt} %
	%	\setlength{\belowcaptionskip}{1pt} % 
        \resizebox{0.48\textwidth}{!}{
	\begin{tabular}{lcccc}
		\hline
		Dataset                                                                                   & of Nodes                                                                  & \begin{tabular}[c]{@{}c@{}}of Edges \\ (in millions)\end{tabular} & \begin{tabular}[c]{@{}c@{}}Attack \\ batch/node\end{tabular}                       & Size (GB)                                                    \\ \hline
		StreamSpot                                                                                & 999,999                                                                   & 89.8                                                              & 100 batches                                                                          & 2.8                                                          \\ \hline
		Unicorn Wget                                                                              & 39,606,900                                                                & 145.9                                                             & 25 batches                                                                           & 76.6                                                         \\ \hline
		\begin{tabular}[c]{@{}l@{}}DARPA E3-TRACE\\ DARPA E3-CADETS\\ DARPA E3-THEIA\end{tabular} & \begin{tabular}[c]{@{}c@{}}3,288,676\\ 1,627,035\\ 1,623,966\end{tabular} & \begin{tabular}[c]{@{}c@{}}4\\ 2.8\\ 3.3\end{tabular}             & \begin{tabular}[c]{@{}c@{}}68,082 nodes\\ 25,319 nodes\\ 12,846 nodes\end{tabular} & \begin{tabular}[c]{@{}c@{}}15\\ 18\\ 18\end{tabular} \\ \hline
	\end{tabular}
	}
	\label{DATASETS}
	
\end{table}

\subsubsection{\bf Baseline Methods}
To compare the efficacy and efficiency of \Slot with state-of-the-art methods, we selected the following seven advanced APT detections at both the graph-level, node-level and edge-level granularity for evaluation. Specialized expert rules or threat reports for APT detection schemes are difficult to compare, so they are not taken into consideration.

\noindent\textbf{1) graph-level detection}
\begin{itemize}[leftmargin=*]
	\item \textbf{StreamSpot}~\cite{manzoor2016fast}: It introduces a new similarity function for graphs, comparing them based on the relative frequency of local substructures represented as short strings.
	% with the consideration of their bias scores.
	\item \textbf{Unicorn}~\cite{han2020unicorn}: UNICORN uses graph sketching to build an incrementally updatable, fixed size, longitudinal graph data structure 1 that enables efficient computation of graph statistics.
	
\end{itemize}

\noindent\textbf{2) node-level detection}
\begin{itemize}[leftmargin=*]
	\item \textbf{Threatrace}~\cite{wang2022threatrace}: It uses GraphSAGE to execute node-level anomaly detection, learning the structural information of nodes and identifying anomalies based on deviations from this learned behavior.
	\item \textbf{Log2vec}~\cite{liu2019log2vec}: It transforms the user's log entries into heterogeneous graphs based on rules and expresses behavioral relationships through graph embedding.
	\item \textbf{FLASH}~\cite{rehman2024flash}: It uses word2vec for semantic learning and GraphSAGE~\cite{hamilton2017inductive} for graph structure to jointly represent log behavior for anomaly detection.
        \item \textbf{MAGIC}~\cite{jia2024magic}: It learns system behavior through a masked autoencoder~\cite{hou2022graphmae} and then uses KNN to calculate node distances, identifying abnormal nodes.
\end{itemize}

\noindent\textbf{3) edge-level detection}
\begin{itemize}[leftmargin=*]
	\item \textbf{ShadeWatcher}~\cite{zengy2022shadewatcher}: It leverages TransR~\cite{lin2015learning} and GNN to detect APTs based on recommendation.
\end{itemize}

\subsubsection{\bf Implementation}
We implement \Slot using Pytorch, comprising approximately 3,000 lines of code. All models operate on Python 3.9, utilizing the Gensim library for the Word2Vec model. Regarding hyperparameters, we set all embedding dimensions to 64, and the learning rate to 0.01. We use Adam as the optimizer, with the similarity loss weight ($\lambda_1$) set at $2$ and the $L2$ regularization weight ($\lambda_2$) set at 0.01. The RL action step size ($\tau$) is configured to 0.02. In Section~\ref{sec:parameter_study}, we provide a detailed discussion on the impact of hyperparameter settings, offering a thorough analysis and explanation of the rationale behind the settings and the experimental logic. All experiments are performed on a server running Ubuntu 18.04 LTS with an Intel 13th Gen Core i7-1360P CPU (12 cores, 16 threads, base frequency 2.20 GHz), 32 GB RAM.

\subsection{\bf Efficacy of Semi-supervision Detection (RQ1)}

\begin{table}[htp]
	\centering
	\caption{\Slot’s detection results on different datasets. For batched log level detection, the detection targets are log pieces. And for system entity level detection, system entities are the targets.}
	\label{tab:EXPERIMENTAL RESULTS}
	\resizebox{\columnwidth}{!}{%
\begin{tabular}{|c|c|cc|cccc|}
	\hline
	\multirow{2}{*}{Granularity}   & \multirow{2}{*}{Dataset} & \multicolumn{2}{c|}{Ground Truth}           & \multicolumn{1}{c|}{\multirow{2}{*}{\#TP}} & \multicolumn{1}{c|}{\multirow{2}{*}{\#FP}} & \multicolumn{1}{c|}{\multirow{2}{*}{\#TN}} & \multirow{2}{*}{\#FN} \\ \cline{3-4}
	&                          & \multicolumn{1}{c|}{\#Benign} & \#Malicious & \multicolumn{1}{c|}{}                      & \multicolumn{1}{c|}{}                      & \multicolumn{1}{c|}{}                      &                       \\ \hline
	\multirow{2}{*}{Batched log}   & StreamSpot               & \multicolumn{1}{c|}{100}      & 90          & 89                                         & 1                                          & 99                                         & 1                     \\ \cline{2-4}
	& Unicorn Wget             & \multicolumn{1}{c|}{25}       & 22          & 21.8                                       & 0.9                                        & 24.1                                       & 0.2                   \\ \cline{1-4}
	\multirow{3}{*}{System entity} & DARPA E3 Trace           & \multicolumn{1}{c|}{613,050}  & 60,644      & 60,604                                     & 1,521                                      & 611,529                                    & 40                    \\ \cline{2-4}
	& DARPA E3 CADETS          & \multicolumn{1}{c|}{343,428}  & 11,561      & 11,550                                     & 478                                        & 342,950                                    & 11                    \\ \cline{2-4}
	& DARPA E3 THEIA           & \multicolumn{1}{c|}{315,212}  & 22,825      & 22,822                                     & 400                                        & 314,812                                    & 3                     \\ \hline
\end{tabular}%
	}
\end{table}

Slot performs node-level fine-grained APT detection on batch logs and event-level logs. In order to reasonably perform result statistics at the batch processing level, we calculate the detection performance of Slot based on the proportion of malicious attacks in the batch processing. Slot uses only 10\% of the malignant structures through oversampling to provide supervision signals for similarity labels and achieve semi-supervised learning. Table~\ref{tab:EXPERIMENTAL RESULTS} shows that Slot accurately detected APT attacks in various scenarios. A small number of malicious structures effectively assist the model in learning and characterizing benign patterns. The node-level Slot achieves very low false negatives and false positives in fine-grained detection. 

\begin{table}[]
\centering
\caption{Impact of Malignant Structure Percentage in Batch and Event-Level Training Datasets on Detection Efficacy.}
\label{tab:train-percentage}
\resizebox{\columnwidth}{!}{%
\begin{tabular}{cccccccl}
\cline{1-7}
\multicolumn{1}{|c|}{Dateset}                     & \multicolumn{1}{c|}{Train\%} & \multicolumn{1}{c|}{Accuracy} & \multicolumn{1}{c|}{Precison} & \multicolumn{1}{c|}{Recall} & \multicolumn{1}{c|}{F1-Score} & \multicolumn{1}{c|}{AUC}    &  \\ \cline{1-7}
\multicolumn{1}{|c|}{\multirow{4}{*}{StreamSpot}} & \multicolumn{1}{c|}{5}       & \multicolumn{1}{c|}{0.8924}   & \multicolumn{1}{c|}{0.8922}   & \multicolumn{1}{c|}{0.9115} & \multicolumn{1}{c|}{0.8911}   & \multicolumn{1}{c|}{0.9213} &  \\ \cline{2-7}
\multicolumn{1}{|c|}{}                            & \multicolumn{1}{c|}{10}      & \multicolumn{1}{c|}{0.9999}   & \multicolumn{1}{c|}{0.9999}   & \multicolumn{1}{c|}{0.9999} & \multicolumn{1}{c|}{0.9999}   & \multicolumn{1}{c|}{0.9999} &  \\ \cline{2-7}
\multicolumn{1}{|c|}{}                            & \multicolumn{1}{c|}{20}      & \multicolumn{1}{c|}{0.9999}   & \multicolumn{1}{c|}{0.9999}   & \multicolumn{1}{c|}{0.9999} & \multicolumn{1}{c|}{0.9999}   & \multicolumn{1}{c|}{0.9999} &  \\ \cline{2-7}
\multicolumn{1}{|c|}{}                            & \multicolumn{1}{c|}{40}      & \multicolumn{1}{c|}{0.9999}   & \multicolumn{1}{c|}{0.9999}   & \multicolumn{1}{c|}{0.9999} & \multicolumn{1}{c|}{0.9999}   & \multicolumn{1}{c|}{0.9999} &  \\ \cline{1-7}
\multicolumn{1}{|c|}{\multirow{4}{*}{CADETS}}     & \multicolumn{1}{c|}{5}       & \multicolumn{1}{c|}{0.9069}   & \multicolumn{1}{c|}{0.9016}   & \multicolumn{1}{c|}{0.9203} & \multicolumn{1}{c|}{0.9050}   & \multicolumn{1}{c|}{0.9780} &  \\ \cline{2-7}
\multicolumn{1}{|c|}{}                            & \multicolumn{1}{c|}{10}      & \multicolumn{1}{c|}{0.9986}   & \multicolumn{1}{c|}{0.9602}   & \multicolumn{1}{c|}{0.9990} & \multicolumn{1}{c|}{0.9792}   & \multicolumn{1}{c|}{0.9970} &  \\ \cline{2-7}
\multicolumn{1}{|c|}{}                            & \multicolumn{1}{c|}{20}      & \multicolumn{1}{c|}{0.9907}   & \multicolumn{1}{c|}{0.9895}   & \multicolumn{1}{c|}{0.9911} & \multicolumn{1}{c|}{0.9903}   & \multicolumn{1}{c|}{0.9981} &  \\ \cline{2-7}
\multicolumn{1}{|c|}{}                            & \multicolumn{1}{c|}{40}      & \multicolumn{1}{c|}{0.9999}   & \multicolumn{1}{c|}{0.9999}   & \multicolumn{1}{c|}{0.9999} & \multicolumn{1}{c|}{0.9999}   & \multicolumn{1}{c|}{0.9999} &  \\ \cline{1-7}
\multicolumn{1}{l}{}                              & \multicolumn{1}{l}{}         & \multicolumn{1}{l}{}          & \multicolumn{1}{l}{}          & \multicolumn{1}{l}{}        & \multicolumn{1}{l}{}          & \multicolumn{1}{l}{}        & 
\end{tabular}%
}
\end{table}

Through similarity awareness, Slot can effectively select neighbors that are more similar to the central node, allowing the Graph Neural Network (GNN) to better capture patterns in system behavior. Table \ref{tab:train-percentage} demonstrates the advantages of semi-supervised learning. While a greater number of malicious signals lead to significant improvements in performance, the results show that even a small amount of malicious supervision signals is sufficient to train an effective model.

\subsection{Overall Detection Efficacy Comparsion (RQ2)}

\begin{table}[htbp]
	\caption{\revision{The comparison results between \Slot and state-of-the-art approaches.}}
	\label{tab:COMPARISON}
	\centering
	\tabcolsep=2pt
  \resizebox{0.48\textwidth}{!}{
	%\footnotesize
        \begin{threeparttable}

\begin{tabular}{@{}ccccccc@{}}
	\toprule
	Datasets                                                  & System                                                                                              & Accuracy                                                                                                                                                                                                         & Precison                                                                                                                                                                                                         & Recall                                                                                                                                                                                                                & F1-Score                                                                                                                                                                                                         & FPR                                                                                            \\ \midrule
	StreamSpot                                                & \begin{tabular}[c]{@{}c@{}}StreamSpot\\ Unicorn\\ Threatrace\\ \Slot\end{tabular}                   & \begin{tabular}[c]{@{}c@{}}$93\%_{\dec{6.06}}$\\ $99\%_{\dec{0.00}}$\\ $99\%_{\dec{0.00}}$\\ \textbf{99\%}$_{\inc{2.06}}$\end{tabular}                                                                           & \begin{tabular}[c]{@{}c@{}}$73\%_{\dec{26.2}}$\\ $95\%_{\dec{4.04}}$\\ $98\%_{\dec{1.01}}$\\ \textbf{99\%}$_{\inc{11.7}}$\end{tabular}                                                                           & \begin{tabular}[c]{@{}c@{}}$91\%_{\dec{8.08}}$\\ $97\%_{\dec{2.02}}$\\ $99\%_{\dec{0.00}}$\\ \textbf{99\%}$_{\inc{3.48}}$\end{tabular}                                                                                & \begin{tabular}[c]{@{}c@{}}$81\%_{\dec{18.2}}$\\ $96\%_{\dec{3.03}}$\\ $99\%_{\dec{0.00}}$\\ \textbf{99\%}$_{\inc{7.60}}$\end{tabular}                                                                           & \begin{tabular}[c]{@{}c@{}}$6.6\%_{\inc{3200}}$\\  $1.6\%_{\inc{700}}$\\ $0.4\%_{\inc{100}}$\\ \textbf{0.2\%}$_{\dec{93.0}}$\end{tabular}                 \\ \midrule
	\begin{tabular}[c]{@{}c@{}}Unicorn\\ Wget\end{tabular}    & \begin{tabular}[c]{@{}c@{}}Unicorn\\ Threatrace\\ \Slot\end{tabular}                                & \begin{tabular}[c]{@{}c@{}}$90\%_{\dec{9.09}}$\\ $95\%_{\dec{4.04}}$\\ \textbf{99\%}$_{\inc{7.02}}$\end{tabular}                                                                                                 & \begin{tabular}[c]{@{}c@{}}$86\%_{\dec{10.4}}$\\ $93\%_{\dec{3.12}}$\\ \textbf{96\%}$_{\inc{7.26}}$\end{tabular}                                                                                                 & \begin{tabular}[c]{@{}c@{}}$95\%_{\dec{4.04}}$\\ $98\%_{\dec{1.01}}$\\ \textbf{99\%}$_{\inc{2.59}}$\end{tabular}                                                                                                      & \begin{tabular}[c]{@{}c@{}}$90\%_{\dec{7.21}}$\\ $95\%_{\dec{2.06}}$\\ \textbf{97\%}$_{\inc{4.86}}$\end{tabular}                                                                                                 & \begin{tabular}[c]{@{}c@{}}$15.5\%_{\inc{330}}$\\ $7.4\%_{\inc{105}}$\\ \textbf{3.6\%}$_{\dec{68.5}}$\end{tabular}                        \\ \midrule
	\begin{tabular}[c]{@{}c@{}}DARPA E3\\ TRACE\end{tabular}  & \begin{tabular}[c]{@{}c@{}}Log2vec\\ ThreaTrace\\ ShadeWatcher\\ FLASH\\ MAGIC\\ \Slot\end{tabular} & \begin{tabular}[c]{@{}c@{}}$97.63\%_{\dec{2.13}}$\\ $98.92\%_{\dec{0.84}}$\\ \textbf{99.96\%}$_{\inc{0.20}}$\\ $99.75\%_{\dec{0.01}}$\\ $99.04\%_{\dec{0.72}}$\\ \underline{99.76\%}$_{\inc{0.70}}$\end{tabular} & \begin{tabular}[c]{@{}c@{}}$54.41\%_{\dec{44.2}}$\\ $71.59\%_{\dec{26.6}}$\\ \textbf{99.98\%}$_{\inc{2.49}}$\\ $94.66\%_{\dec{2.96}}$\\ $92.00\%_{\dec{5.68}}$\\ \underline{97.55\%}$_{\inc{18.2}}$\end{tabular} & \begin{tabular}[c]{@{}c@{}}$78.25\%_{\dec{21.6}}$\\ $99.99\%_{\inc{0.06}}$\\ \textbf{99.99\%}$_{\inc{0.06}}$\\ \textbf{99.99\%}$_{\inc{0.06}}$\\ $98.99\%_{\dec{0.94}}$\\ \underline{99.93\%}$_{\inc{4.70}}$\end{tabular} & \begin{tabular}[c]{@{}c@{}}$64.18\%_{\dec{34.9}}$\\ $83.43\%_{\dec{15.4}}$\\ \textbf{99.98\%}$_{\inc{1.27}}$\\ $97.20\%_{\dec{1.53}}$\\ $95.43\%_{\dec{3.33}}$\\ \underline{$98.72\%_{\inc{12.1}}$}\end{tabular} & \begin{tabular}[c]{@{}c@{}}$1.8\%_{\inc{800}}$\\ $1.1\%_{\inc{450}}$\\ $0.3\%_{\inc{50}}$\\ $0.3\%_{\inc{50}}$\\ $0.9\%_{\inc{350}}$\\ \textbf{0.2\%}$_{\dec{77.2}}$\end{tabular} \\ \midrule
	\begin{tabular}[c]{@{}c@{}}DARPA E3\\ CADETS\end{tabular} & \begin{tabular}[c]{@{}c@{}}Log2vec\\ ThreaTrace\\ FLASH\\ MAGIC\\ \Slot\end{tabular}                & \begin{tabular}[c]{@{}c@{}}$98.16\%_{\dec{1.70}}$\\ $99.81\%_{\dec{0.05}}$\\ $99.73\%_{\dec{0.13}}$\\ $99.18\%_{\dec{0.68}}$\\ \textbf{99.86\%}$_{\inc{0.64}}$\end{tabular}                                      & \begin{tabular}[c]{@{}c@{}}$49.20\%_{\dec{48.7}}$\\ $90.42\%_{\dec{5.83}}$\\ $93.43\%_{\dec{2.69}}$\\ $82.02\%_{\dec{14.5}}$\\ \textbf{96.02\%}$_{\inc{21.8}}$\end{tabular}                                      & \begin{tabular}[c]{@{}c@{}}$84.55\%_{\dec{15.3}}$\\ $99.97\%_{\inc{0.07}}$\\ $99.94\%_{\inc{0.04}}$\\ $99.07\%_{\dec{0.83}}$\\ \textbf{99.90\%}$_{\inc{4.19}}$\end{tabular}                                           & \begin{tabular}[c]{@{}c@{}}$62.23\%_{\dec{36.4}}$\\ $94.95\%_{\dec{3.03}}$\\ $96.58\%_{\dec{1.36}}$\\ $89.75\%_{\dec{8.34}}$\\ \textbf{97.92\%}$_{\inc{14.0}}$\end{tabular}                                      & \begin{tabular}[c]{@{}c@{}}$1.6\%_{\inc{1600}}$\\ $0.2\%_{\inc{100}}$\\ $0.3\%_{\inc{200}}$\\ $0.8\%_{\inc{700}}$\\ \textbf{0.1\%}$_{\dec{86.2}}$\end{tabular}         \\ \midrule
	\begin{tabular}[c]{@{}c@{}}DARPA E3\\ THEIA\end{tabular}  & \begin{tabular}[c]{@{}c@{}}Log2vec\\ ThreaTrace\\ FLASH\\ MAGIC\\ \Slot\end{tabular}                & \begin{tabular}[c]{@{}c@{}}$99.55\%_{\dec{0.33}}$\\ $99.87\%_{\dec{0.01}}$\\ $99.38\%_{\dec{0.50}}$\\ $99.86\%_{\dec{0.02}}$\\ \textbf{99.88\%}$_{\inc{0.22}}$\end{tabular}                                      & \begin{tabular}[c]{@{}c@{}}$62.49\%_{\dec{36.4}}$\\ $87.03\%_{\dec{11.4}}$\\ $91.73\%_{\dec{6.65}}$\\ $98.23\%_{\dec{0.04}}$\\ \textbf{98.27\%}$_{\inc{15.7}}$\end{tabular}                                      & \begin{tabular}[c]{@{}c@{}}$66.05\%_{\dec{33.9}}$\\ $99.74\%_{\dec{0.24}}$\\ $99.82\%_{\dec{0.16}}$\\ $99.99\%_{\dec{0.00}}$\\ \textbf{99.99\%}$_{\inc{9.39}}$\end{tabular}                                           & \begin{tabular}[c]{@{}c@{}}$64.23\%_{\dec{35.2}}$\\ $92.95\%_{\dec{6.23}}$\\ $95.63\%_{\dec{3.53}}$\\ $99.10\%_{\dec{0.03}}$\\ \textbf{99.13\%}$_{\inc{12.6}}$\end{tabular}                                      & \begin{tabular}[c]{@{}c@{}}$0.3\%_{\inc{150}}$\\ \textbf{0.1\%}$_{\dec{16.6}}$\\ $0.8\%_{\inc{566}}$\\ $0.14\%_{\inc{16.6}}$\\ \underline{0.12\%}$_{\dec{64.2}}$\end{tabular}         \\ \bottomrule
\end{tabular}%

\begin{tablenotes}[para,flushleft]
\item[] \textbf{Bold} denotes the best results, and \underline{$underlined$} denotes the second-best results.
\item[] $\dec{}$ represents the percentage of decrease, $\inc{}$ represents the percentage of improvement.
\item[1] \revision{Improvement compared to the average of all detection systems except \Slot.}

\item[2] \revision{FPR means False Positive Rate. A detection system with a lower false positive rate can better avoid false positive attacks during detection.}

\end{tablenotes}
\end{threeparttable}

}
\end{table}

The effectiveness of \Slot was tested using both batch and entity-level logs. To demonstrate \Slot's precise ability to distinguish between benign and malicious behaviors, a fair comparison was made with relevant SOTA (state-of-the-art) methods, as shown in Table~\ref{tab:COMPARISON}. Based on these results, we draw the following observations:

%\Slot was tested for effectiveness using both batch and entity-level logs. Table~\ref{tab:EXPERIMENTAL RESULTS} shows that \Slot accurately detected APT attacks across various scenarios. From these results, we make the following observations:

\begin{itemize}[leftmargin=*]
\item \textbf{Obs.1}: \textbf{Comparison in Batch Logs}. For batch audit log detection, we used the datasets provided by Streamspot and Unicorn as baselines and compared the results with node-level detection methods, including Threatrace. In the relatively simple Streamspot logs, where attack patterns are straightforward and behavioral differences are clear, all detection methods performed well. However, \Slot achieved near-perfect detection results. On the Unicorn dataset, where highly camouflaged attacks significantly reduced the detection efficacy of other methods, \Slot excelled by distinguishing node similarities and strengthening relationship selection. This effectively magnified the differences between benign and malicious behaviors, resulting in superior detection efficacy.

\item \textbf{Obs.2}: \textbf{Comparison in Entity-Level Logs}. In entity-level log detection, we compared \Slot with other node-level methods. Log2vec treats logs as nodes rather than entities, fragmenting entity information and weakening aggregation. GNN-based methods like MAGIC, Threatrace, and FLASH also failed to address benign behaviors camouflaging malicious entities, making them vulnerable to evasion. In contrast, \Slot used reinforcement learning to filter out benign-camouflaged neighbors, resulting in superior detection efficacy compared to the SOTA node-level system, FLASH. In Figure 6, compared to FLASH, the \Slot ROC curve demonstrates superior model classification performance.

\end{itemize}

\subsection{System Overhead (RQ3)}
\begin{table}[htp]
	\centering
	\caption{The time cost and proportion of each stage for \Slot on the training graph with 300K nodes.}
	\label{tab:time of RL}
	\resizebox{\columnwidth}{!}{%
		\begin{tabular}{@{}|l|l|l|l|@{}}
			\toprule
			Phase      & Graph  Construction & Reinforcement Learning & GNN     \\ \midrule
			Time cost  & 46.12s              & 331.94s                & 64.69s  \\ \midrule
			Percentage & 10.41\%             & 74.97\%                & 14.67\% \\ \bottomrule
		\end{tabular}%
	}
\end{table}
\begin{figure}[htp]
    \centering
    \subfigure[Detection rates with changes in batch size.]{
		\includegraphics[width=1.47in]{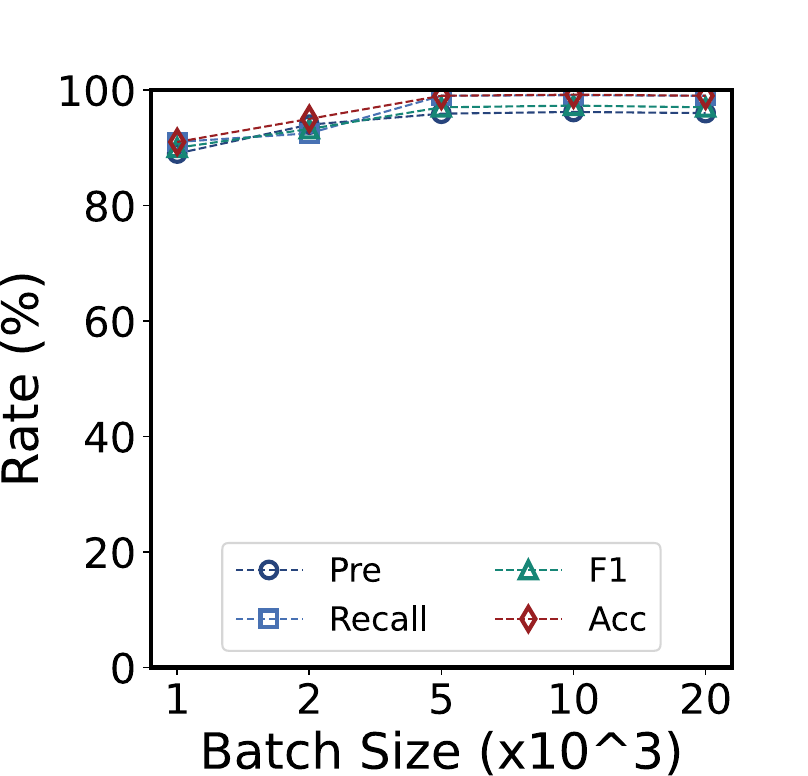}
        \label{fig: Detection effectiveness}
    }
    \subfigure[RAM and CPU overheads with changes in batch size.]{
        \includegraphics[width=1.53in]{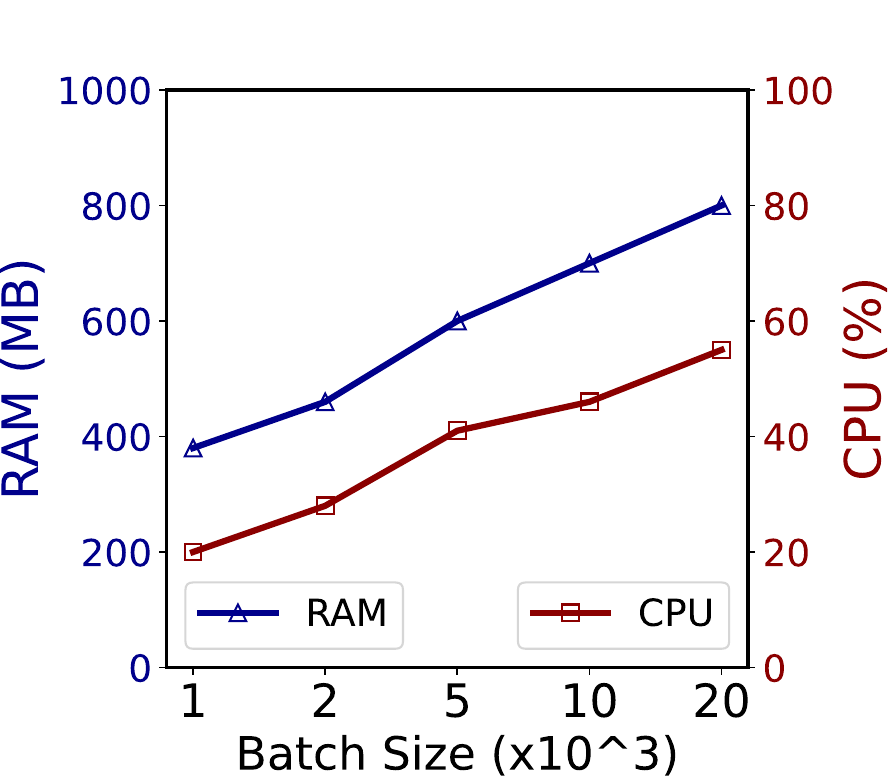}
        \label{fig: RAM and CPU}
    }
    \caption{Detection efficacy and efficiency of \Slot across various batch sizes.}
\end{figure}

%\Slot是一种基于GNN的检测方案，因此在检测推理阶段的性能开销取决于所构图的大小，由batch size参数决定。我们通过调整batch size来全面的分析了\Slot关于CPU、内存、时间的开销变化。

% 图7a 展示了batchsize对检测精度影响，因为图神经网络处理时，更大的图批次有利于学习到更为完整的局部结构，随着批次增大，检测精度也更优，但当超过某个阈值(完成系统大多数行为所需的最大事件实体数)后，检测能力变得稳定。考虑到batchsize对检测精度有着重要影响，我们分析了在不同batchsize下CPU、和内存的开销。图7b中展示了，更大的batchsize会提供更加复杂的图结构和节点信息，在图信息传播过程需要更大的内存和节点计算。同样图7c展示，随着每批次处理的节点增多，时间呈线性增长。

\Slot is a detection scheme based on GNN, and therefore, the efficiency overhead during the detection phase depends on the size of the constructed graph, which is determined by the batch size parameter. We comprehensively analyzed the changes in efficiency, CPU and memory overhead of \Slot by adjusting the batch size. And compared the time overhead when selecting a batch size of 5000.

Figure~\ref{fig: Detection effectiveness} shows the impact of batch size on detection effectiveness. Since larger graph batches facilitate learning a more complete local structure in GNN processing, the detection efficiency improves as the batch size increases. However, beyond a certain threshold (The maximum number of entities required to complete most of the system activities), the detection capability stabilizes. Considering that batch size has a significant impact on detection accuracy, we analyzed the CPU and memory overhead under different batch sizes. As shown in Figure~\ref{fig: RAM and CPU}, larger batch sizes provide more complex graph structures and node information, which requires more memory and node computation during the graph information propagation process. In terms of detection time, while the introduction of reinforcement learning incurs additional overhead, as shown in Table~\ref{tab:time of RL}, our tests on graphs with 300K nodes demonstrate that although reinforcement learning consumes more time, the overhead can be considered acceptable.

%In terms of detection time, as shown in Figure~\ref{fig: time with batch size}, a larger batch size corresponds to increased detection time. We observe that when the batch size exceeds 5000, the performance improvement in detection becomes marginal, as 5000 nodes are sufficient to ensure that most system behaviors are not fragmented. Under the condition of a batch size of 5000, Figure~\ref{fig: time} illustrates that \Slot achieves faster detection speed compared to MAGIC~\cite{jia2024magic}. Although MAGIC enhances the effectiveness of GNN-based anomaly detection using masking techniques, its KNN distance calculations for anomaly detection still require a significant amount of time.

\subsection{Module Ablation Study (RQ4)}

\begin{figure}[htp]
	\centering

	\includegraphics[width=\columnwidth]{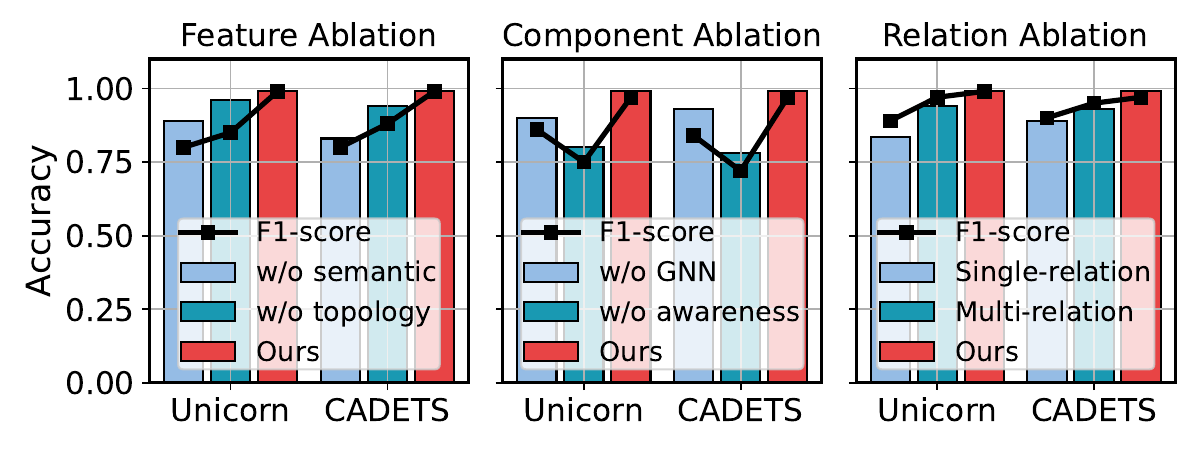}
	\caption{The impact on accuracy and F1 score from separately ablating features, components, and relations, evaluated on the Unicorn and CADETS datasets.}

	\label{fig:ablation}
\end{figure}

In our ablation experiments, we re-evaluate the efficacy of \Slot by varying the content of various components to show their impact on the effectiveness of the system. The results are reported in Figure~\ref{fig:ablation}.

\noindent \textbf{Effect of semantic features and topological features}. In the similarity-aware module, we conducted ablation studies on semantic and topological structures to explore the value of node semantic information and node topology in the selection of node neighbors. Experimental results show that both have a certain impact on node neighbor aggregation. During feature embedding, integrating semantic features and structural features simultaneously can effectively distinguish between benign and malicious nodes. This approach significantly improves the accuracy of similarity segmentation.

\noindent \textbf{Effect of similarity awareness \& GNN}. \Slot characterizes node features through similarity awareness and GNN. We conducted ablation tests on the main modules of \Slot, and experiments revealed that using similarity awareness alone cannot deeply explore graph structure information, while GNN without the guidance of neighbor selection is susceptible to the influence of noisy neighbors, misleading node embeddings. Experimental results demonstrate that GNN guided by similarity significantly enhances detection efficacy.

\noindent \textbf{Effect of provenance relation}.
Current APT detection~\cite{wang2022threatrace, rehman2024flash, zengy2022shadewatcher} based on GNN often treat provenance graphs as homogeneous, failing to differentiate the heterogeneity of node neighbors, with all relationships being merged into the central node. As illustrated in Figure~\ref{fig:ablation}, we compared the detection efficacy of three methods: homogeneous aggregation, multi-relation aggregation, and deep relational aggregation based on hidden relation mining. The experimental results indicate that neighbor relationship filtering based on reinforced selection effectively reduces the induced expression of benign neighbors on malicious nodes. Additionally, by incorporating hidden relational information, the model can mine deeper feature information, thereby accelerating the effective propagation of information flow and enhancing detection efficiency.

\begin{figure}[htp]
	\centering
	\includegraphics[width=0.47\columnwidth]{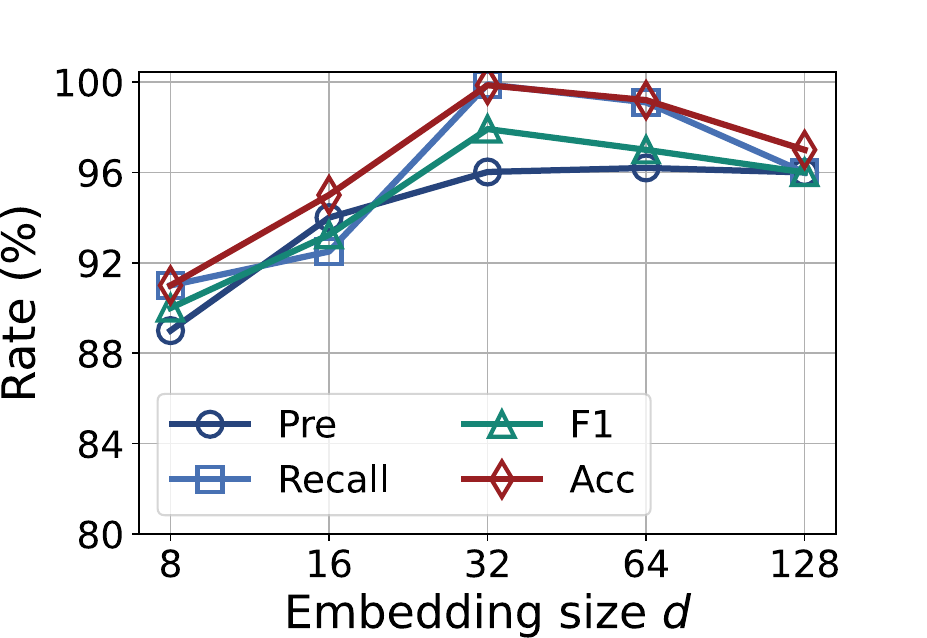}
	\includegraphics[width=0.49\columnwidth]{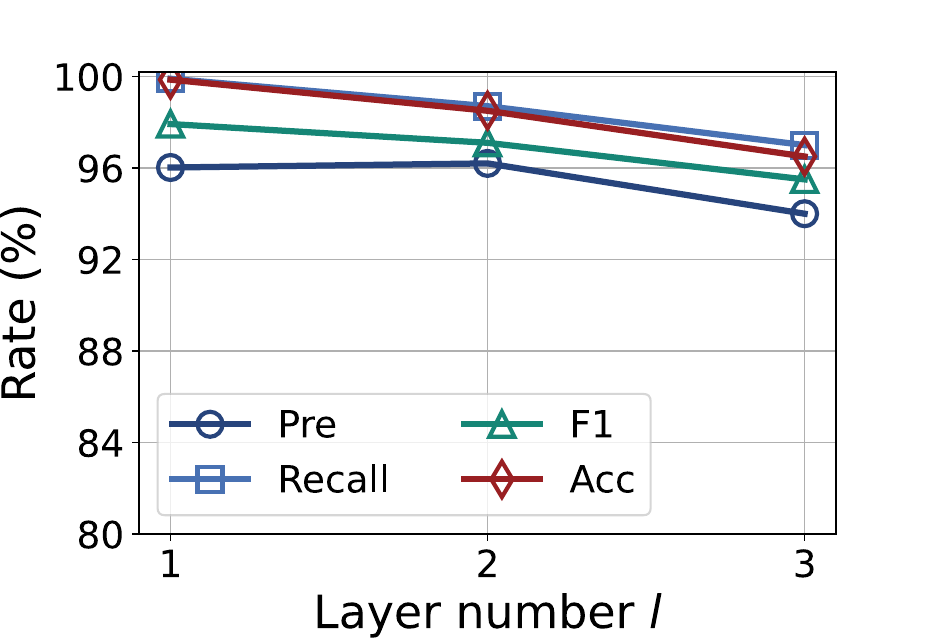}
	\includegraphics[width=0.49\columnwidth]{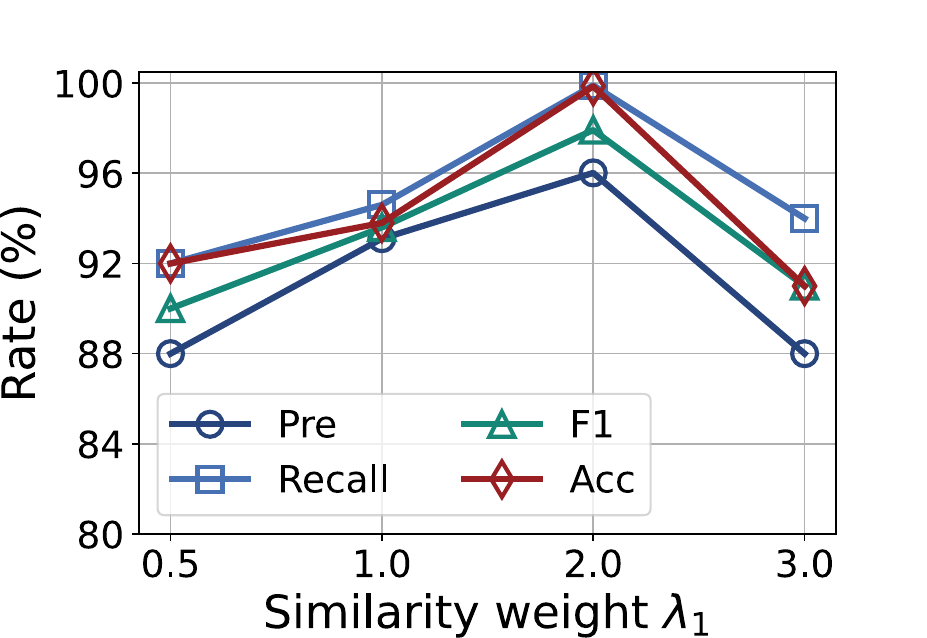}
	\includegraphics[width=0.49\columnwidth]{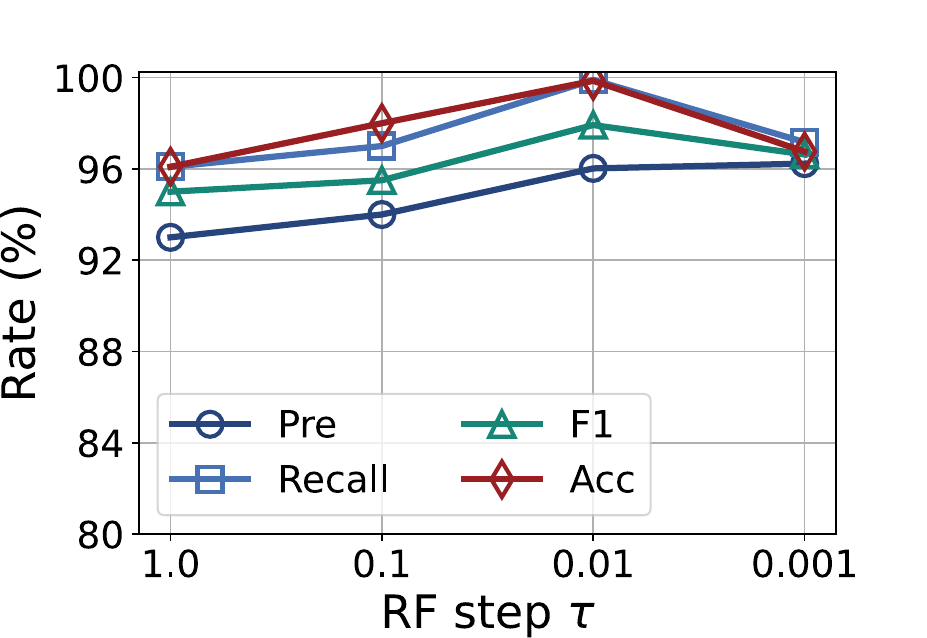}
	\caption{Detection efficacy (i.e., accuracy, precision, recall, F1) with different hyperparameters on the CADETS dataset.}
	\label{fig:Hyperparameter}
	
\end{figure} 

\begin{figure*}[htp]
	\centering
	\subfigure[System Activities Near Backdoor Attack Events.]{
		\includegraphics[width=1.85in, height=1.34in]{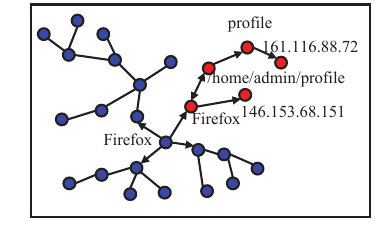}
		\label{fig: locakattack}
	}
	\subfigure[Node Embedding of System Events Near Backdoor Attack Activities Detected by FLASH.]{
		\includegraphics[width=2in]{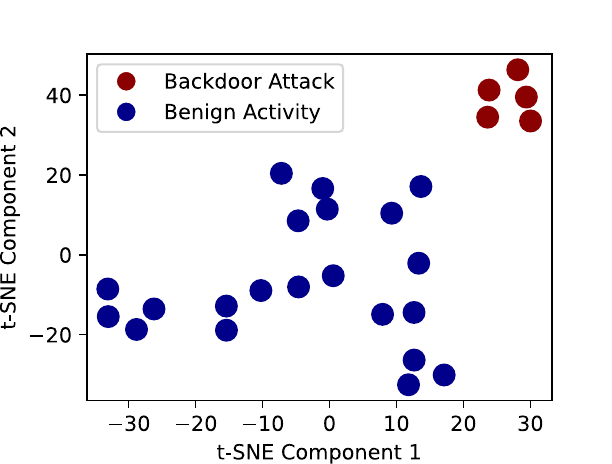}
		\label{fig: flash-t-SNE-1}
	}
	\subfigure[Node Embedding of System Events Near Backdoor Attack Activities Detected by \Slot.]{
		\includegraphics[width=2in]{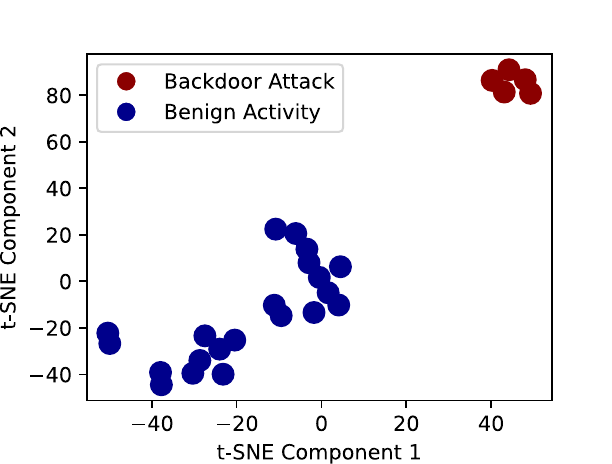}   
		\label{fig: Slot-t-SNE-1}
	}
	\subfigure[System Event Structure After Cloning Multiple Benign DNS Resolution Activities at Attack Nodes.]{
		\centering
		\includegraphics[width=1.85in, height=1.33in]{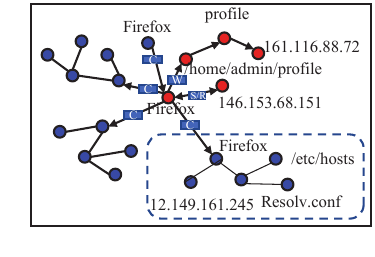}
		\label{fig: Adversarial Attacks Event Structure}
	}
	\subfigure[Node Embedding of System Events Near Backdoor Attack Activities Detected by FLASH under Adversarial Attacks.]{
		\includegraphics[width=2in]{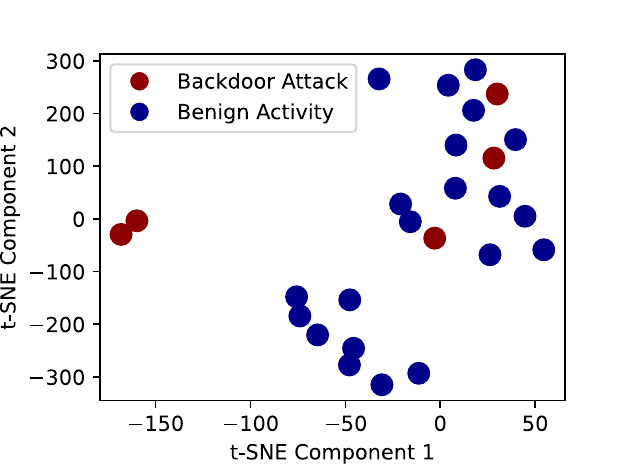}
		\label{fig: flash-t-SNE-2}
	}
	\subfigure[Node Embedding of System Events Near Backdoor Attack Activities Detected by \Slot under Adversarial Attacks]{
		\includegraphics[width=2in]{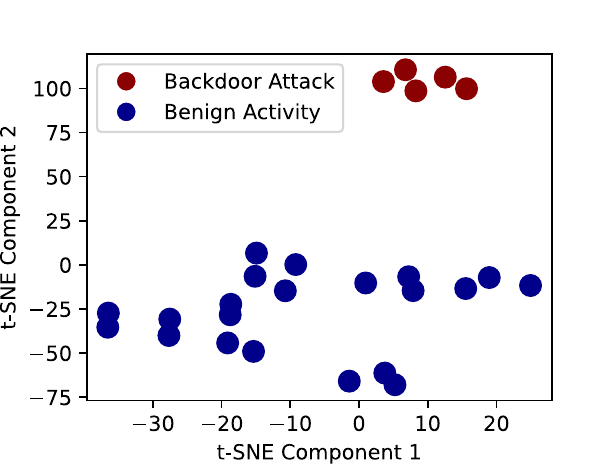}   
		\label{fig: Slot-t-SNE-2}
	}
	\caption{t-SNE Visualization of Adversarial Attack Embeddings.}
\end{figure*}

\subsection{Hyperparameter Investigation (RQ5)}

\label{sec:parameter_study}
Previously, we employed a fixed set of optimal parameters; in this section, we discuss the impact of several critical hyperparameters on the threat detection efficacy of \Slot. We examined the logic behind the choice of hyperparameters separately on CADETS shown in Figure~\ref{fig:Hyperparameter}. When testing one of these parameters, we maintained the others consistent with the baseline. We summarize the results with the following observations:

\begin{itemize}[leftmargin=*]

    \item \textbf{Embedding Size} $d$: In the embedding process, node features are embedded and mapped to low-dimensional vectors while preserving as much of their original feature attributes as possible. As shown in Figure~\ref{fig:Hyperparameter}, as the embedding dimension increases, the detection metrics improve significantly, indicating that a larger embedding dimension effectively captures more information. However, an excessively large dimension can lead to feature sparsity, which severely impacts detection efficacy and results in higher memory consumption. When the dimension is set to 32, the model achieves the fastest convergence and reaches high efficacy metrics.
    
     \item \textbf{Number of Layers} $l$: Generally, increasing the number of layers in a model can capture deeper information. However, as shown in Figure~\ref{fig:Hyperparameter}, there is no significant improvement in efficacy with an increase in the number of layers. Particularly, when the model exceeds three layers, there is a clear occurrence of overfitting. Therefore, choosing a single-layer model can achieve the best balance between detection efficacy and efficiency cost.
    
    \item \textbf{Similarity Loss Weight} $\lambda_{1}$: Figure~\ref{fig:ablation} demonstrates the significant impact of similarity calculations on \Slot. Figure~\ref{fig:Hyperparameter} shows the effect of different similarity loss weights on detection efficacy. It is evident that when the weight of the similarity loss is twice that of the GNN loss, there is a balanced integration of feature similarity perception and GNN information aggregation, resulting in optimal efficacy.
    
    \item \textbf{Reinforcement Learning step} $\tau$: 
    In reinforcement learning, selecting the appropriate action step size allows all thresholds to be updated and converge over multiple epochs. When the thresholds oscillate for several rounds, reaching Formula~\ref{terminal} is the termination condition. Experiments have demonstrated that \Slot achieve optimal efficacy when the step size is set to 0.01. 

\end{itemize}

\subsection{Resilience to Adversarial Attacks (RQ6)}
\label{sec:Resilience to Adversarial Attacks}

\begin{figure}[htp]
	\centering
	\includegraphics[width=0.48\textwidth]{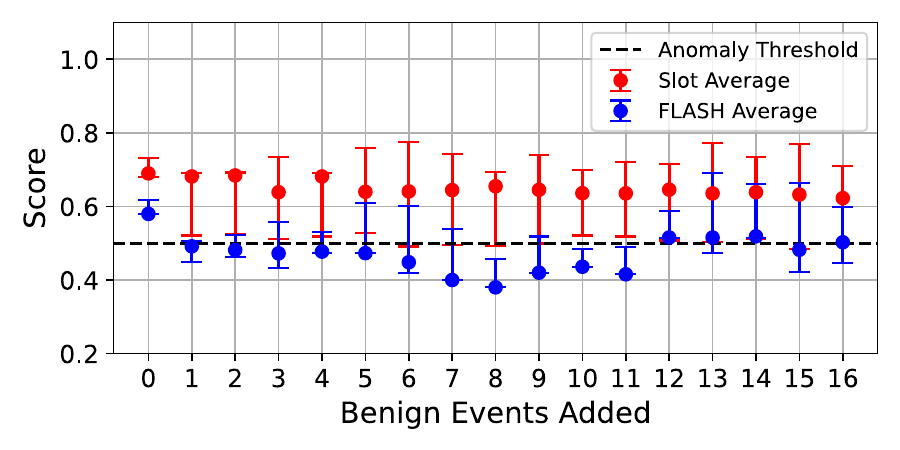}
	
	\caption{Resilience against adversarial attacks.}
	\label{fig:adversarial attack}
	
\end{figure}

\noindent{\bf Adversarial Attack}:
Mimicry attacks against provenance-based APT detection are carried out by altering the original data to incorporate more benign features, thereby 'mimicking benign behavior' in order to evade detection. To evaluate the resilience of the \Slot system against adversarial attacks, we employed the attack methods described in~\cite{goyal2023sometimes} and~\cite{rehman2024flash}, which involve inserting benign structures into the attack graph. This allowed us to assess the robustness of \Slot.
% 为了展示Slot对对抗性攻击的鲁棒性，我们以THEIA dataset为例，来说明Slot如何能区分模仿性的良性结构与恶性行为之间的差异，以防止攻击方通过模仿良性行为来诱导模型对恶性节点产生错误表达。 图10a为firefox后门漏洞利用中攻击活动附近的系统行为，攻击活动与良性行为有着较少的交互关系，因此图10b中仅通过GNN刻画可以使良性行为和攻击行为有相对明显的嵌入划分，而图10c中Slot通过半监督的相似性标签，可以使得不同性质的节点在嵌入空间中有着更大的区分距离。 图10d中，我们将虚线框中常见的良性行为(DNS解析)通过Clone方式，在不影响攻击活动的情况下，多次插入了攻击节点附近，来干扰模型对攻击活动的嵌入判断。 图10e可以看出，受良性模仿结构影响，与良性结构靠近的攻击行为会被表达为良性。 而图10f中，Slot通过相似性感知，可以对firefox的邻居动态选择不同的关系聚合权重，从而可以免受大量的Clone操作带来的良性结构误导。因此仍表现出较好的嵌入区分。

To demonstrate the robustness of Slot against adversarial attacks, we use the THEIA dataset as an example to show how Slot can distinguish between benign imitative structures and malicious behaviors, preventing attackers from misleading the model by mimicking benign behaviors and inducing incorrect representations of malicious nodes. In Figure~\ref{fig: locakattack}, we present the system behavior near the exploitation of a Firefox backdoor vulnerability. The attack activity has minimal interaction with benign behaviors, allowing for clear separation between benign and attack behaviors when using GNN embeddings, as shown in Figure~\ref{fig: flash-t-SNE-1}. In contrast, Figure~\ref{fig: Slot-t-SNE-1} illustrates that Slot, by leveraging semi-supervised similarity labels, enables a greater distinction between different types of nodes in the embedding space. In Figure~\ref{fig: Adversarial Attacks Event Structure}, we use the Clone technique to repeatedly insert common benign behaviors (DNS resolution) near attack nodes without affecting the attack activity, in an attempt to interfere with the model's embedding judgment of the attack. As seen in Figure~\ref{fig: flash-t-SNE-2}, the attack behaviors, influenced by benign imitative structures, are mistakenly represented as benign when they are close to benign behaviors. However, as shown in Figure~\ref{fig: Slot-t-SNE-2}, Slot counters this by adapting its aggregation weights based on neighborhood similarity, allowing it to selectively aggregate different types of relationships in Firefox's neighborhood. This mechanism enables the model to resist the misleading effects of multiple Clone operations, resulting in a better separation of the embeddings despite the presence of benign structure imitations.

As shown in Figure~\ref{fig:adversarial attack}, both \Slot and FLASH exhibit a certain level of robustness against imitation attacks when only a small number of benign events are introduced. However, as more benign events are incorporated, we observe a significant decline in FLASH's detection efficacy, while \Slot remains unaffected by the imitation attack, demonstrating greater robustness compared to the baseline. We attribute this to FLASH and other graph-based representation learning methods for APT detection, which fail to adequately consider the heterogeneous relationships between nodes during propagation and aggregation. A small number of benign embeddings have minimal impact at the node level, but when more benign structures aggregate near attack nodes, it leads to erroneous representations of the attack nodes. In contrast, \Slot effectively guides neighbor selection by distinguishing behavioral similarities, making it resilient to the degree of adversarial influence. By employing multi-armed bandit-based reinforcement learning, \Slot dynamically filters out benign camouflage related to malicious nodes, demonstrating strong resistance against imitation attacks.

% \vspace{-0.1in}
\subsection{Alert Validation (RQ7)}

\begin{figure}[htp]
	\centering
	\includegraphics[width=0.45\textwidth]{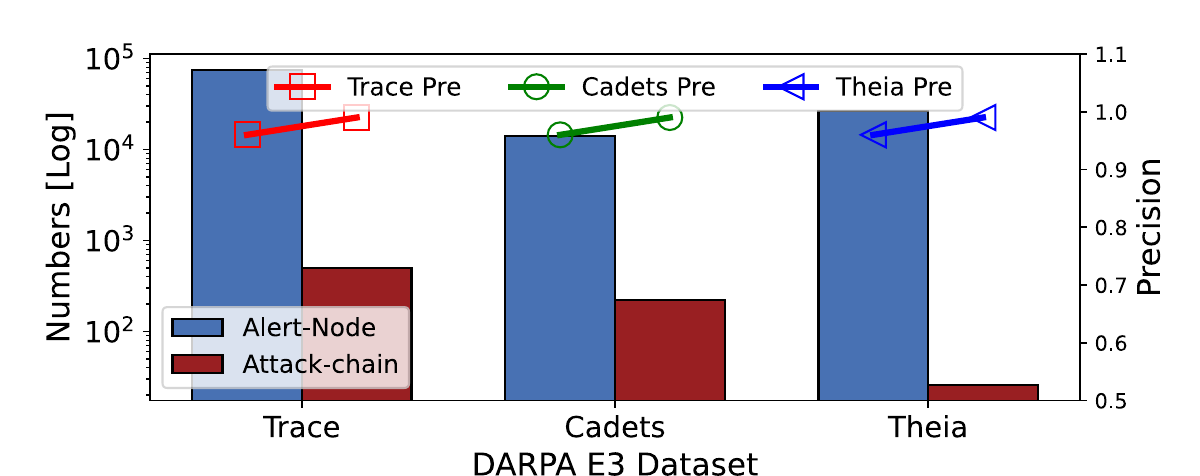}

	\caption{The number of alerts produced with the DARPA E3 dataset and the false positive rate within the generated attack chain.}
	\label{fig:Alert Validation}

\end{figure}

The ultimate goal of \Slot is to provide security analysts with more effective and concise attack insights, thereby accelerating the alert-handling process. We believe that APT detection should achieve the following when reconstructing attack chains: (1) Provide concise and readable attack chains, (2) Include as few false positive nodes in the attack chain as possible.

\begin{figure}[htp]
	\centering
	\includegraphics[width=0.35\textwidth]{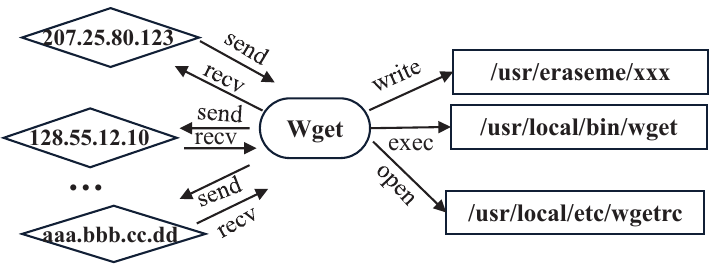}
	\caption{An example of benign activities reconstructed within the alert nodes.}
	\label{fig:Benign activity}

\end{figure}

\begin{figure}[htp]
	\centering
	\includegraphics[width=0.35\textwidth]{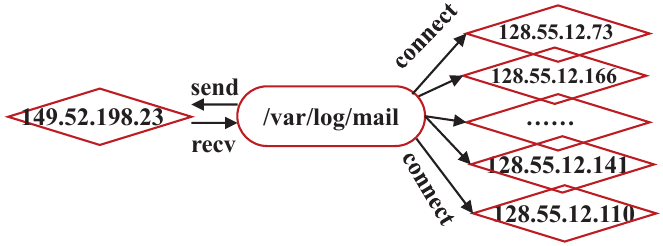}
	\caption{An example of attack activities reconstructed within the alert nodes.}
	\label{fig:attack activity}

\end{figure}
%As shown in Figure~\ref{fig:Alert Validation}, \Slot can rapidly filter and identify the most relevant paths during the complex attack chain decision-making process, using TTP to exclude paths that may appear related but are commonly found in normal operations. This further enhances the filtering of false positives in attack-chains. Additionally, as described by FLASH~\cite{rehman2024flash}, the reconstructed attack graph exhibits strong event representation capabilities, eliminating the need for security analysts to trace through numerous anomalous nodes. As shown in Figure~\ref{fig:Benign activity}, the benign structure in CADETS, which is a series of activities generated by "$wget$", is highly aggregated. This effectively accelerates the filtering and selection of benign entity graphs. As shown in Figure~\ref{fig:attack activity}, the attack scanning behavior in THEIA initiated by the malicious process "$/var/log/email$" involves more than 6000 scanned network entities. The aggregation of the attack chain significantly reduces the time required for analysts to manually trace these activities. The use of attack chains significantly reduces the number of items that analysts need to review, as the number of attack chains is substantially smaller than that of anomalous nodes.

As shown in Figure~\ref{fig:Alert Validation}, \Slot rapidly filters and identifies the most relevant paths during the attack chain decision-making process, using TTP to exclude paths that may appear related but are common in normal operations. This enhances false positive filtering in attack chains. Additionally, as described by FLASH~\cite{rehman2024flash}, the reconstructed attack graph exhibits strong event representation, eliminating the need for security analysts to trace numerous anomalous nodes. As shown in Figure~\ref{fig:Benign activity}, the benign structure in CADETS, generated by "$wget$", is highly aggregated, accelerating the filtering and selection of benign entity graphs. In Figure~\ref{fig:attack activity}, the attack scanning behavior in THEIA initiated by the malicious process "$/var/log/email$" involves over 6000 scanned network entities. Aggregating the attack chain reduces the time analysts need to manually trace these activities. The use of attack chains significantly reduces the items analysts need to review, as they are far fewer than anomalous nodes.

%如图12给出的良性结构，由wget产生的一系列活动，他们高度的聚合在了一起。This effectively accelerates the filtering and selection of benign entity graphs.图13所示的攻击扫描行为，包含了6000个以上的被扫描的网络实体，攻击链聚合有效的减少了分析师手动溯源所需要的时间开销。 The use of attack-chain significantly reduces the number of items that analysts need to review, as the number of attack-chains is substantially smaller than that of anomalous nodes. 

\noindent{\bf Case Study}. Using the attack scenario from Section~\ref{sec: Attack}, we demonstrate how \Slot detects APT attacks from audit logs and reconstructs the attack chain. In this scenario, the attacker leverages a Firefox backdoor and embeds benign DNS resolutions to mask their activities. \Slot employs semantic and topological embeddings to derive primary node representations and calculate similarities, distinguishing between benign behaviors (e.g., Firefox accessing common IPs) and malicious activities. By segmenting similarities between Firefox1 and Firefox2 and using reinforcement learning, \Slot filters out benign activities disguised within attack events. Using GNN, it learns the genuine behavioral patterns of nodes. In the final Attack-chain Reconstruction phase, TTP encoding further separates benign peripheral nodes, resulting in the red-highlighted attack chain. Core malicious activities include communication with IP $146.153.68.151$, downloading and executing a malicious file ($/home/admin/profile$), and interacting with IP $149.52.198.23$ to manipulate processes for attack scanning.

%我们使用Section~\ref{sec: Attack}中的攻击场景，来详细说明\Slot如何从审计日志中发现APT攻击并还原攻击场景。在攻击场景中攻击者利用Firefox后门发起了一系列攻击，并且为了伪装行为，在攻击活动中嵌入了用户常见的DNS解析。在检测中，\Slot 通过语义嵌入和拓扑嵌入，学习节点的初级嵌入表达，然后计算其相似性，有效的学习到虚线框中的用户常见良性行为（FireFox对常用ip访问并解析域名）与后门攻击行为之间的差异，从而在FireFox1和FireFox2的相似性计算中有效的实现了分割，然后通过强化学习选择过滤了攻击事件中伪装的良性行为，通过GNN学习了真实的节点行为模式。最后通过Attack-chain Reconstruction，\Slot通过引入TTP编码，进一步分离了边缘良性节点，从而生成图中的红色攻击简图，其核心攻击活动包含了从Firefox后门开始，与恶性ip 146.153.68.151进行通信，然后下载profile恶意文件并执行，以及与恶性ip地址149.52.198.23通信并操纵恶意进程mail实现攻击扫描。

%\noindent{\bf Case Study}. 我们使用第二节中的攻击场景，来详细说明\Slot如何从审计日志中发现APT攻击并还原攻击场景。动机中的攻击场景包含了用户的良性行为和攻击者的攻击行为，首先\Slot 通过语义嵌入和拓扑嵌入，学习节点的初级嵌入表达，然后计算其相似性，\Slot 可以有效的学习到虚线框中的用户常见良性行为（FireFox对常用ip访问并解析域名），从而在FireFox1和FireFox2的相似性计算中有效的实现了分割，然后通过强化学习选择过滤了攻击事件中伪装的良性行为，成功将攻击实体检测了出来，最后通过Alert Reconstruction，\Slot 在攻击链溯源构建中成功建立了攻击故事，生成了更易于分析的Alert-chain。

\subsection{Discussion}
\noindent \textbf{Unknown Attacks.} 
\Slot uses graph reinforcement learning to model both benign and malicious behaviors. However, APT attacks involve numerous unknown 0-day attacks. Although \Slot provides considerations for detecting unknown anomalies, relying solely on feature isolation for anomaly detection tends to result in a high false positive rate, and requires appropriate confidence levels to balance model-based learning detection and anomaly detection. In the future, we aim to extend unsupervised graph reinforcement learning to APT detection.

\noindent \textbf{Concept drift.} 
As user business evolves, system behaviors also change accordingly, leading to new behaviors that the model has not previously encountered. This causes the behavior patterns learned by the model to become outdated. Such concept drift may result in misclassification by the model. Similar to existing strategies~\cite{jia2024magic, zengy2022shadewatcher}, \Slot can also adopt current small-batch incremental learning methods~\cite{tian2024survey} and historical data forgetting mechanisms~\cite{han2020unicorn} to adapt to these changes, thereby enhancing the adaptability and accuracy of the model.

\noindent \textbf{Adversarial attacks.} 
In Section~\ref{sec:Resilience to Adversarial Attacks}, we consider mimicry adversarial attack, which is a form of black-box attack. Adversarial attacks also include gradient-based white-box attacks, but these require the attacker to have full access to the target model, including knowledge of the model architecture, weights, and training data, which is often impractical in real-world scenarios. Moreover, existing methods such as adversarial training~\cite{tramer2019adversarial} and gradient masking~\cite{mkadry2017towards} have shown to be effective defenses, and \Slot can integrate and extend these methods accordingly.

\section{Conclusion}
\label{sec:conclusoin}
% The persistent and evolving threat posed by advanced persistent threats (APTs) necessitates innovative detection solutions that can effectively navigate the complexities of modern cyberattacks. Existing methods often fall short in addressing the full spectrum of challenges associated with APT detection, such as limited detection efficacy, vulnerability to adversarial attacks, and insufficient support in developing defense strategies. To bridge these gaps, we introduced \Slot, an effective APT detection approach that leverages provenance graphs and graph reinforcement learning. With advanced provenance graph digging techniques, \Slot excels in uncovering multi-level hidden relationships among system behaviors. Besides, \Slot leverages graph reinforcement learning to dynamically adapt to new user activities and attack strategies, significantly enhancing its accuracy and resilience in highly adversarial environments. By automatically constructing attack chains, \Slot not only identifies attack paths with precision but also supports the development of effective defense strategies. Our evaluations reveal \Slot's effectiveness, achieving approximately 99\% APT detection accuracy and outperforming state-of-the-art approaches. Further experiments also highlight \Slot's high efficiency, resilience to adversarial attacks, and ability to support the development of defense strategies.
The persistent and evolving threat posed by advanced persistent threats (APTs) necessitates innovative detection solutions that can navigate the complexities of modern cyberattacks. Existing methods often fall short in addressing challenges like limited detection efficacy, vulnerability to adversarial attacks, and insufficient support for developing defense strategies. To bridge these gaps, we introduced \Slot, an APT detection approach leveraging provenance graphs and graph reinforcement learning. \revision{Using advanced provenance graph digging, \Slot uncovers multi-level hidden relationships among system behaviors. \Slot boosts semi-supervised learning with limited labels via efficient label similarity computation, enhancing both detection accuracy and model robustness. Additionally, it leverages graph reinforcement learning to adapt to new user activities and attack strategies, improving accuracy and resilience in adversarial environments.} By automatically constructing attack chains, \Slot identifies attack paths with precision and supports defense strategy development. Our evaluations show \Slot's effectiveness, achieving 99\% APT detection accuracy and outperforming state-of-the-art approaches. Further experiments highlight \Slot's high efficiency, resilience to adversarial attacks, and ability to support defense strategy development.

\section{Acknowledgments}
This research is funded by the National Key Research and Development Program of China (2023YFE0111100), Natural Science Basic Research Program of Shaanxi (No. 2025JC-JCQN-073), National Natural Science Foundation of China under Grant (No. 62272370), Young Elite Scientists Sponsorship Program by CAST (2022QNRC001),  the China 111Project (No.B16037), Qinchuangyuan Scientist + Engineer Team Program of Shaanxi (No. 2024QCY-KXJ-149), Songshan Laboratory (No. 241110210200), Open Foundation of Key Laboratory of Cyberspace Security, Ministry of Education of China (No.KLCS20240405 ) and the Fundamental Research Funds for the Central Universities (QTZX23071), the National Research Foundation, Singapore, and DSO National Laboratories under the AI Singapore Programme (AISG Award No: AISG2-GC-2023-008), the National Research Foundation, Singapore, and the Cyber Security Agency under its National Cybersecurity R\&D Programme (NCRP25-P04-TAICeN), the National Research Foundation, Prime Minister’s Office, Singapore under its Campus for Research Excellence and Technological Enterprise (CREATE) programme, and Ripple under its University Blockchain Research Initiative (UBRI)~\cite{feng2022university}.

\appendix \section{Appendix}
\label{sec:appendix}
\subsection{The multi-class classification performance of StreamSpot.}
\begin{figure}[htp]
	\centering
	\includegraphics[width=0.48\textwidth]{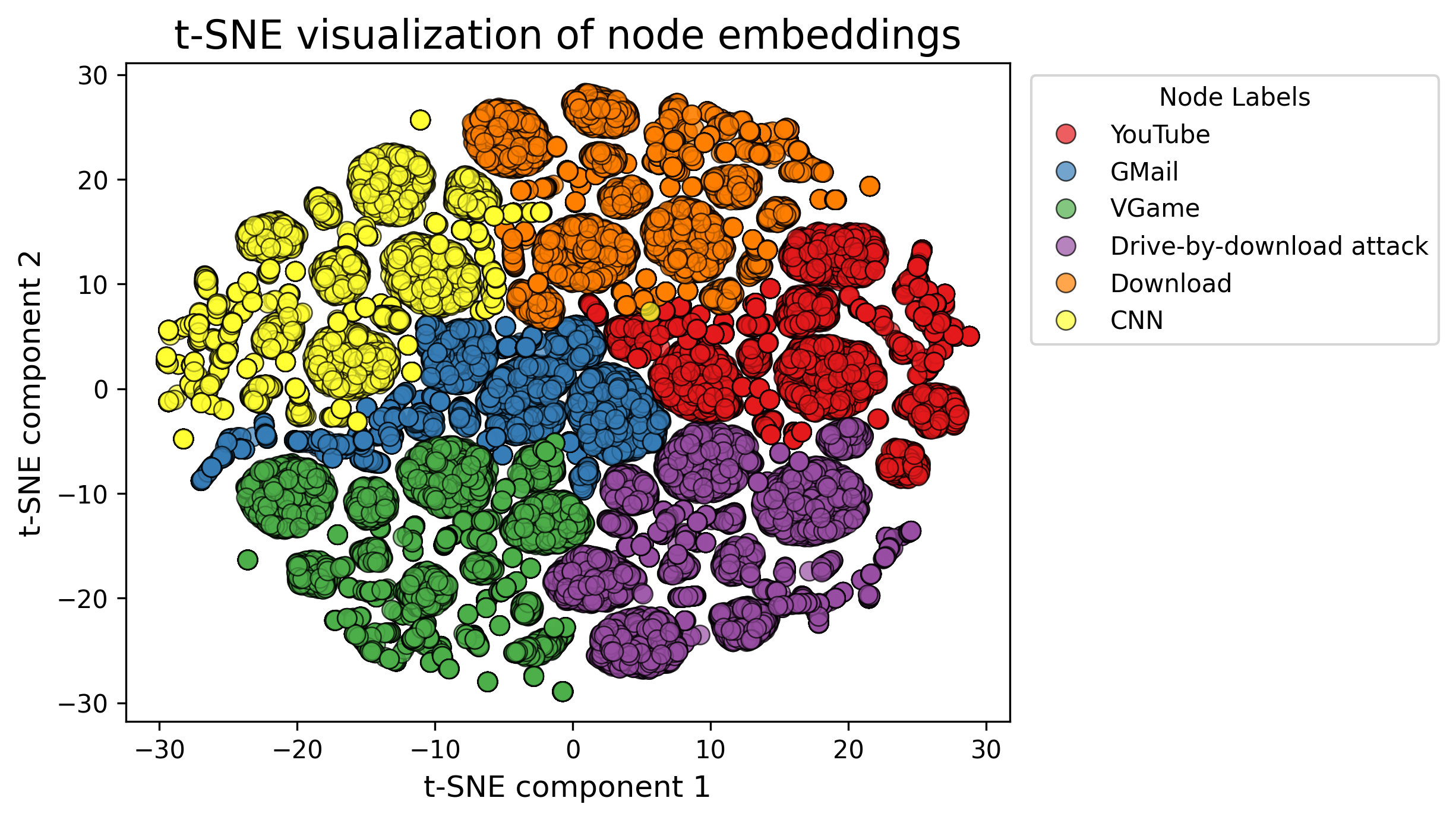}
	
	\caption {Visualization of t-SNE embeddings in the StreamSpot dataset. The dataset includes five user behavior scenarios and one attack scenario: watching YouTube, checking Gmail, playing VGame, experiencing drive-by-download attack, downloading regular files, and watching CNN. Each point represents a system call behavior from the dataset.}
	\label{fig:streamspot-tsne}
\end{figure}

\Slot achieves near-perfect attack detection in the StreamSpot dataset. However, since the StreamSpot dataset also contains ground truth for benign activities, I have similarly performed multi-class classification on the node events. As shown in Figure~\ref{fig:streamspot-tsne}, the six behavior classes exhibit clear clustering, which strongly demonstrates \Slot's powerful capability in similarity awareness for multi-class recognition.

\subsection{Impact of Oversampling Ratio}
% APT攻击日志是一个严重不平衡的数据，正样本数量远小于负样本，模型可能会倾向于预测负类，因为这样能够更好地符合数据中的多数类别。通过过采样正样本，可以增加正样本在训练数据中的比例，从而迫使模型学习如何区分两类，而不是仅仅记住负类。图5实验证明，Slot通过1：1的正样本负样本采样学习，可以实现更好的检测性能。

\begin{figure}[htp]
	\centering
	\includegraphics[width=0.48\textwidth]{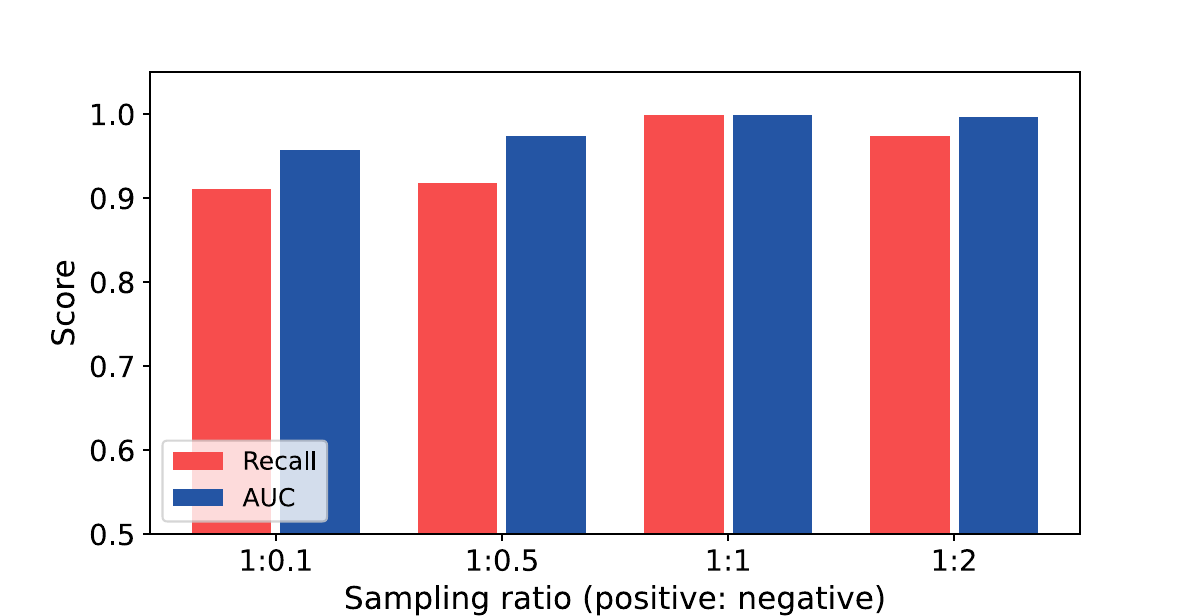}
	
	\caption {The best AUC and Recall results at different sampling ratios over 30 epochs.}
	\label{fig:Oversampling}
\end{figure}

APT attack logs are a severely imbalanced dataset, with the number of positive samples being much smaller than the number of negative samples. The model may tend to predict the negative class because it better fits the majority class in the data. By oversampling the positive samples, the proportion of positive samples in the training data can be increased, thereby forcing the model to learn how to distinguish between the two classes, rather than simply memorizing the negative class. We chose Recall and AUC to measure the impact of sampling ratios on model detection. The reason is that Recall measures the proportion of actual positive samples that are correctly predicted as positive. For oversampling the positive class, Recall reflects how many positive samples the model can identify. AUC measures the model's ability to distinguish between positive and negative samples.The experiment shown in Figure~\ref{fig:Oversampling} demonstrates that Slot, with a 1:1 positive-to-negative sampling ratio, achieves better detection performance.

\bibliographystyle{ACM-Reference-Format}
\balance
\bibliography{sample-base}


\begin{thebibliography}{65}

%%% the Web address.

\ifx \showCODEN    \undefined \def \showCODEN     #1{\unskip}     \fi
\ifx \showISBNx    \undefined \def \showISBNx     #1{\unskip}     \fi
\ifx \showISBNxiii \undefined \def \showISBNxiii  #1{\unskip}     \fi
\ifx \showISSN     \undefined \def \showISSN      #1{\unskip}     \fi
\ifx \showLCCN     \undefined \def \showLCCN      #1{\unskip}     \fi
\ifx \shownote     \undefined \def \shownote      #1{#1}          \fi
\ifx \showarticletitle \undefined \def \showarticletitle #1{#1}   \fi
\ifx \showURL      \undefined \def \showURL       {\relax}        \fi
% invisible to TeX
\providecommand\bibfield[2]{#2}
\providecommand\bibinfo[2]{#2}
\providecommand\natexlab[1]{#1}
\providecommand\showeprint[2][]{arXiv:#2}

\bibitem[DAR({[n.\,d.]})]%
        {DARPA}
 \bibinfo{year}{[n.\,d.]}\natexlab{}.
\newblock \bibinfo{title}{Darpa transparent computing program engagement 3 data
  release}.
\newblock \bibinfo{howpublished}{\url{https://github.com/darpa-i2o/
  Transparent-Computing}}.
\newblock
\newblock
\shownote{2020}.


\bibitem[lin({[n.\,d.]})]%
        {linux_auditd}
 \bibinfo{year}{[n.\,d.]}\natexlab{}.
\newblock \bibinfo{title}{The Linux audit daemon}.
\newblock \bibinfo{howpublished}{\url{https://linux.die.net/man/8/auditd}}.
\newblock


\bibitem[mit({[n.\,d.]})]%
        {mitre_attack}
 \bibinfo{year}{[n.\,d.]}\natexlab{}.
\newblock \bibinfo{title}{{MITRE ATT\&CK Framework}}.
\newblock
\urldef\tempurl%
\url{https://attack.mitre.org/}
\showURL{%
\tempurl}


\bibitem[The({[n.\,d.]})]%
        {Thestreamspotdataset}
 \bibinfo{year}{[n.\,d.]}\natexlab{}.
\newblock \bibinfo{title}{The streamspot dataset}.
\newblock \bibinfo{howpublished}{\url{https://github.com/
  sbustreamspot/sbustreamspot-data}}.
\newblock
\newblock
\shownote{2016}.


\bibitem[Ahmad et~al\mbox{.}(2022)]%
        {ahmad2022hardlog}
\bibfield{author}{\bibinfo{person}{Adil Ahmad}, \bibinfo{person}{Sangho Lee},
  {and} \bibinfo{person}{Marcus Peinado}.} \bibinfo{year}{2022}\natexlab{}.
\newblock \showarticletitle{Hardlog: Practical tamper-proof system auditing
  using a novel audit device}. In \bibinfo{booktitle}{\emph{2022 IEEE Symposium
  on Security and Privacy (SP)}}. IEEE, \bibinfo{pages}{1791--1807}.
\newblock


\bibitem[Alsaheel et~al\mbox{.}(2021)]%
        {alsaheel2021atlas}
\bibfield{author}{\bibinfo{person}{Abdulellah Alsaheel},
  \bibinfo{person}{Yuhong Nan}, \bibinfo{person}{Shiqing Ma},
  \bibinfo{person}{Le Yu}, \bibinfo{person}{Gregory Walkup},
  \bibinfo{person}{Z~Berkay Celik}, \bibinfo{person}{Xiangyu Zhang}, {and}
  \bibinfo{person}{Dongyan Xu}.} \bibinfo{year}{2021}\natexlab{}.
\newblock \showarticletitle{$\{$ATLAS$\}$: A sequence-based learning approach
  for attack investigation}. In \bibinfo{booktitle}{\emph{30th USENIX security
  symposium (USENIX security 21)}}. \bibinfo{pages}{3005--3022}.
\newblock


\bibitem[Alshamrani et~al\mbox{.}(2019)]%
        {alshamrani2019survey}
\bibfield{author}{\bibinfo{person}{Adel Alshamrani}, \bibinfo{person}{Sowmya
  Myneni}, \bibinfo{person}{Ankur Chowdhary}, {and} \bibinfo{person}{Dijiang
  Huang}.} \bibinfo{year}{2019}\natexlab{}.
\newblock \showarticletitle{A survey on advanced persistent threats:
  Techniques, solutions, challenges, and research opportunities}.
\newblock \bibinfo{journal}{\emph{IEEE Communications Surveys \& Tutorials}}
  \bibinfo{volume}{21}, \bibinfo{number}{2} (\bibinfo{year}{2019}),
  \bibinfo{pages}{1851--1877}.
\newblock


\bibitem[Bacciu and Numeroso(2022)]%
        {bacciu2022explaining}
\bibfield{author}{\bibinfo{person}{Davide Bacciu} {and} \bibinfo{person}{Danilo
  Numeroso}.} \bibinfo{year}{2022}\natexlab{}.
\newblock \showarticletitle{Explaining deep graph networks via input
  perturbation}.
\newblock \bibinfo{journal}{\emph{IEEE Transactions on Neural Networks and
  Learning Systems}} \bibinfo{volume}{34}, \bibinfo{number}{12}
  (\bibinfo{year}{2022}), \bibinfo{pages}{10334--10345}.
\newblock


\bibitem[Chen et~al\mbox{.}(2022)]%
        {chen2022game}
\bibfield{author}{\bibinfo{person}{Miaojiang Chen}, \bibinfo{person}{Wei Liu},
  \bibinfo{person}{Tian Wang}, \bibinfo{person}{Shaobo Zhang}, {and}
  \bibinfo{person}{Anfeng Liu}.} \bibinfo{year}{2022}\natexlab{}.
\newblock \showarticletitle{A game-based deep reinforcement learning approach
  for energy-efficient computation in MEC systems}.
\newblock \bibinfo{journal}{\emph{Knowledge-Based Systems}}
  \bibinfo{volume}{235} (\bibinfo{year}{2022}), \bibinfo{pages}{107660}.
\newblock


\bibitem[Cheng et~al\mbox{.}(2023)]%
        {cheng2023kairos}
\bibfield{author}{\bibinfo{person}{Zijun Cheng}, \bibinfo{person}{Qiujian Lv},
  \bibinfo{person}{Jinyuan Liang}, \bibinfo{person}{Yan Wang},
  \bibinfo{person}{Degang Sun}, \bibinfo{person}{Thomas Pasquier}, {and}
  \bibinfo{person}{Xueyuan Han}.} \bibinfo{year}{2023}\natexlab{}.
\newblock \showarticletitle{Kairos:: Practical Intrusion Detection and
  Investigation using Whole-system Provenance}.
\newblock \bibinfo{journal}{\emph{arXiv preprint arXiv:2308.05034}}
  (\bibinfo{year}{2023}).
\newblock


\bibitem[Cheng et~al\mbox{.}(2019)]%
        {cheng2019outlier}
\bibfield{author}{\bibinfo{person}{Zhangyu Cheng}, \bibinfo{person}{Chengming
  Zou}, {and} \bibinfo{person}{Jianwei Dong}.} \bibinfo{year}{2019}\natexlab{}.
\newblock \showarticletitle{Outlier detection using isolation forest and local
  outlier factor}. In \bibinfo{booktitle}{\emph{Proceedings of the conference
  on research in adaptive and convergent systems}}. \bibinfo{pages}{161--168}.
\newblock


\bibitem[Corporation({[n.\,d.]})]%
        {microsoft_event_tracing}
\bibfield{author}{\bibinfo{person}{Microsoft Corporation}.}
  \bibinfo{year}{[n.\,d.]}\natexlab{}.
\newblock \bibinfo{title}{Event tracing}.
\newblock
  \bibinfo{howpublished}{\url{https://docs.microsoft.com/en-us/windows/desktop/ETW/event-tracing-portal}}.
\newblock


\bibitem[Feng et~al\mbox{.}(2025)]%
        {feng2025unmasking}
\bibfield{author}{\bibinfo{person}{Yebo Feng}, \bibinfo{person}{Jun Li},
  \bibinfo{person}{Jelena Mirkovic}, \bibinfo{person}{Cong Wu},
  \bibinfo{person}{Chong Wang}, \bibinfo{person}{Hao Ren},
  \bibinfo{person}{Jiahua Xu}, {and} \bibinfo{person}{Yang Liu}.}
  \bibinfo{year}{2025}\natexlab{}.
\newblock \showarticletitle{Unmasking the Internet: A Survey of Fine-Grained
  Network Traffic Analysis}.
\newblock \bibinfo{journal}{\emph{IEEE Communications Surveys \& Tutorials}}
  (\bibinfo{year}{2025}).
\newblock


\bibitem[Feng et~al\mbox{.}(2020)]%
        {feng2020application}
\bibfield{author}{\bibinfo{person}{Yebo Feng}, \bibinfo{person}{Jun Li}, {and}
  \bibinfo{person}{Thanh Nguyen}.} \bibinfo{year}{2020}\natexlab{}.
\newblock \showarticletitle{Application-layer DDoS defense with reinforcement
  learning}. In \bibinfo{booktitle}{\emph{2020 IEEE/ACM 28th International
  Symposium on Quality of Service (IWQoS)}}. IEEE, \bibinfo{pages}{1--10}.
\newblock


\bibitem[Feng et~al\mbox{.}(2023)]%
        {feng2023explainable}
\bibfield{author}{\bibinfo{person}{Yebo Feng}, \bibinfo{person}{Jun Li},
  \bibinfo{person}{Devkishen Sisodia}, {and} \bibinfo{person}{Peter Reiher}.}
  \bibinfo{year}{2023}\natexlab{}.
\newblock \showarticletitle{On explainable and adaptable detection of
  distributed denial-of-service traffic}.
\newblock \bibinfo{journal}{\emph{IEEE Transactions on Dependable and Secure
  Computing}} \bibinfo{volume}{21}, \bibinfo{number}{4} (\bibinfo{year}{2023}),
  \bibinfo{pages}{2211--2226}.
\newblock


\bibitem[Feng et~al\mbox{.}(2022)]%
        {feng2022university}
\bibfield{author}{\bibinfo{person}{Yebo Feng}, \bibinfo{person}{Jiahua Xu},
  {and} \bibinfo{person}{Lauren Weymouth}.} \bibinfo{year}{2022}\natexlab{}.
\newblock \showarticletitle{University blockchain research initiative (ubri):
  Boosting blockchain education and research}.
\newblock \bibinfo{journal}{\emph{IEEE Potentials}} \bibinfo{volume}{41},
  \bibinfo{number}{6} (\bibinfo{year}{2022}), \bibinfo{pages}{19--25}.
\newblock


\bibitem[Goyal et~al\mbox{.}(2023)]%
        {goyal2023sometimes}
\bibfield{author}{\bibinfo{person}{Akul Goyal}, \bibinfo{person}{Xueyuan Han},
  \bibinfo{person}{Gang Wang}, {and} \bibinfo{person}{Adam Bates}.}
  \bibinfo{year}{2023}\natexlab{}.
\newblock \showarticletitle{Sometimes, you aren't what you do: Mimicry attacks
  against provenance graph host intrusion detection systems}. In
  \bibinfo{booktitle}{\emph{30th Network and Distributed System Security
  Symposium}}.
\newblock


\bibitem[Gui et~al\mbox{.}(2024)]%
        {gui2024survey}
\bibfield{author}{\bibinfo{person}{Qian Gui}, \bibinfo{person}{Hong Zhou},
  \bibinfo{person}{Na Guo}, {and} \bibinfo{person}{Baoning Niu}.}
  \bibinfo{year}{2024}\natexlab{}.
\newblock \showarticletitle{A survey of class-imbalanced semi-supervised
  learning}.
\newblock \bibinfo{journal}{\emph{Machine Learning}} \bibinfo{volume}{113},
  \bibinfo{number}{8} (\bibinfo{year}{2024}), \bibinfo{pages}{5057--5086}.
\newblock


\bibitem[Hamilton et~al\mbox{.}(2017)]%
        {hamilton2017inductive}
\bibfield{author}{\bibinfo{person}{Will Hamilton}, \bibinfo{person}{Zhitao
  Ying}, {and} \bibinfo{person}{Jure Leskovec}.}
  \bibinfo{year}{2017}\natexlab{}.
\newblock \showarticletitle{Inductive representation learning on large graphs}.
\newblock \bibinfo{journal}{\emph{Advances in neural information processing
  systems}}  \bibinfo{volume}{30} (\bibinfo{year}{2017}).
\newblock


\bibitem[Han et~al\mbox{.}(2020)]%
        {han2020unicorn}
\bibfield{author}{\bibinfo{person}{Xueyuan Han}, \bibinfo{person}{Thomas
  Pasquier}, \bibinfo{person}{Adam Bates}, \bibinfo{person}{James Mickens},
  {and} \bibinfo{person}{Margo Seltzer}.} \bibinfo{year}{2020}\natexlab{}.
\newblock \showarticletitle{UNICORN: Runtime Provenance-Based Detector for
  Advanced Persistent Threats}. In \bibinfo{booktitle}{\emph{27th Annual
  Network and Distributed System Security Symposium, NDSS}}.
\newblock


\bibitem[Hassan et~al\mbox{.}(2020)]%
        {hassan2020tactical}
\bibfield{author}{\bibinfo{person}{Wajih~Ul Hassan}, \bibinfo{person}{Adam
  Bates}, {and} \bibinfo{person}{Daniel Marino}.}
  \bibinfo{year}{2020}\natexlab{}.
\newblock \showarticletitle{Tactical provenance analysis for endpoint detection
  and response systems}. In \bibinfo{booktitle}{\emph{2020 IEEE Symposium on
  Security and Privacy (SP)}}. IEEE, \bibinfo{pages}{1172--1189}.
\newblock


\bibitem[Hassan et~al\mbox{.}(2019)]%
        {hassan2019nodoze}
\bibfield{author}{\bibinfo{person}{Wajih~Ul Hassan}, \bibinfo{person}{Shengjian
  Guo}, \bibinfo{person}{Ding Li}, \bibinfo{person}{Zhengzhang Chen},
  \bibinfo{person}{Kangkook Jee}, \bibinfo{person}{Zhichun Li}, {and}
  \bibinfo{person}{Adam Bates}.} \bibinfo{year}{2019}\natexlab{}.
\newblock \showarticletitle{Nodoze: Combatting threat alert fatigue with
  automated provenance triage}. In \bibinfo{booktitle}{\emph{network and
  distributed systems security symposium}}.
\newblock


\bibitem[Hossain et~al\mbox{.}(2017)]%
        {hossain2017sleuth}
\bibfield{author}{\bibinfo{person}{Md~Nahid Hossain}, \bibinfo{person}{Sadegh~M
  Milajerdi}, \bibinfo{person}{Junao Wang}, \bibinfo{person}{Birhanu Eshete},
  \bibinfo{person}{Rigel Gjomemo}, \bibinfo{person}{R Sekar},
  \bibinfo{person}{Scott Stoller}, {and} \bibinfo{person}{VN Venkatakrishnan}.}
  \bibinfo{year}{2017}\natexlab{}.
\newblock \showarticletitle{SLEUTH: Real-time attack scenario reconstruction
  from COTS audit data}. In \bibinfo{booktitle}{\emph{26th USENIX Security
  Symposium (USENIX Security 17)}}. \bibinfo{pages}{487--504}.
\newblock


\bibitem[Hossain et~al\mbox{.}(2020)]%
        {hossain2020combating}
\bibfield{author}{\bibinfo{person}{Md~Nahid Hossain}, \bibinfo{person}{Sanaz
  Sheikhi}, {and} \bibinfo{person}{R Sekar}.} \bibinfo{year}{2020}\natexlab{}.
\newblock \showarticletitle{Combating dependence explosion in forensic analysis
  using alternative tag propagation semantics}. In
  \bibinfo{booktitle}{\emph{2020 IEEE Symposium on Security and Privacy (SP)}}.
  IEEE, \bibinfo{pages}{1139--1155}.
\newblock


\bibitem[Hou et~al\mbox{.}(2022)]%
        {hou2022graphmae}
\bibfield{author}{\bibinfo{person}{Zhenyu Hou}, \bibinfo{person}{Xiao Liu},
  \bibinfo{person}{Yukuo Cen}, \bibinfo{person}{Yuxiao Dong},
  \bibinfo{person}{Hongxia Yang}, \bibinfo{person}{Chunjie Wang}, {and}
  \bibinfo{person}{Jie Tang}.} \bibinfo{year}{2022}\natexlab{}.
\newblock \showarticletitle{Graphmae: Self-supervised masked graph
  autoencoders}. In \bibinfo{booktitle}{\emph{Proceedings of the 28th ACM
  SIGKDD Conference on Knowledge Discovery and Data Mining}}.
  \bibinfo{pages}{594--604}.
\newblock


\bibitem[Jia et~al\mbox{.}(2024)]%
        {jia2024magic}
\bibfield{author}{\bibinfo{person}{Zian Jia}, \bibinfo{person}{Yun Xiong},
  \bibinfo{person}{Yuhong Nan}, \bibinfo{person}{Yao Zhang},
  \bibinfo{person}{Jinjing Zhao}, {and} \bibinfo{person}{Mi Wen}.}
  \bibinfo{year}{2024}\natexlab{}.
\newblock \showarticletitle{MAGIC: Detecting Advanced Persistent Threats via
  Masked Graph Representation Learning}. In \bibinfo{booktitle}{\emph{33rd
  USENIX Security Symposium (USENIX Security 24)}}.
  \bibinfo{pages}{5197--5214}.
\newblock


\bibitem[Jiang et~al\mbox{.}(2022)]%
        {jiang2022camouflaged}
\bibfield{author}{\bibinfo{person}{Chao Jiang}, \bibinfo{person}{Yi He},
  \bibinfo{person}{Richard Chapman}, {and} \bibinfo{person}{Hongyi Wu}.}
  \bibinfo{year}{2022}\natexlab{}.
\newblock \showarticletitle{Camouflaged poisoning attack on graph neural
  networks}. In \bibinfo{booktitle}{\emph{Proceedings of the 2022 International
  Conference on Multimedia Retrieval}}. \bibinfo{pages}{451--461}.
\newblock


\bibitem[Jiang et~al\mbox{.}(2018)]%
        {jiang2018graph}
\bibfield{author}{\bibinfo{person}{Jiechuan Jiang}, \bibinfo{person}{Chen Dun},
  \bibinfo{person}{Tiejun Huang}, {and} \bibinfo{person}{Zongqing Lu}.}
  \bibinfo{year}{2018}\natexlab{}.
\newblock \showarticletitle{Graph convolutional reinforcement learning}.
\newblock \bibinfo{journal}{\emph{arXiv preprint arXiv:1810.09202}}
  (\bibinfo{year}{2018}).
\newblock


\bibitem[Kapoor et~al\mbox{.}(2021)]%
        {kapoor2021prov}
\bibfield{author}{\bibinfo{person}{Maya Kapoor}, \bibinfo{person}{Joshua
  Melton}, \bibinfo{person}{Michael Ridenhour}, \bibinfo{person}{Siddharth
  Krishnan}, {and} \bibinfo{person}{Thomas Moyer}.}
  \bibinfo{year}{2021}\natexlab{}.
\newblock \showarticletitle{PROV-GEM: automated provenance analysis framework
  using graph embeddings}. In \bibinfo{booktitle}{\emph{2021 20th IEEE
  International Conference on Machine Learning and Applications (ICMLA)}}.
  IEEE, \bibinfo{pages}{1720--1727}.
\newblock


\bibitem[Levine et~al\mbox{.}(2018)]%
        {levine2018learning}
\bibfield{author}{\bibinfo{person}{Sergey Levine}, \bibinfo{person}{Peter
  Pastor}, \bibinfo{person}{Alex Krizhevsky}, \bibinfo{person}{Julian Ibarz},
  {and} \bibinfo{person}{Deirdre Quillen}.} \bibinfo{year}{2018}\natexlab{}.
\newblock \showarticletitle{Learning hand-eye coordination for robotic grasping
  with deep learning and large-scale data collection}.
\newblock \bibinfo{journal}{\emph{The International journal of robotics
  research}} \bibinfo{volume}{37}, \bibinfo{number}{4-5}
  (\bibinfo{year}{2018}), \bibinfo{pages}{421--436}.
\newblock


\bibitem[Lim et~al\mbox{.}(2021)]%
        {lim2021large}
\bibfield{author}{\bibinfo{person}{Derek Lim}, \bibinfo{person}{Felix Hohne},
  \bibinfo{person}{Xiuyu Li}, \bibinfo{person}{Sijia~Linda Huang},
  \bibinfo{person}{Vaishnavi Gupta}, \bibinfo{person}{Omkar Bhalerao}, {and}
  \bibinfo{person}{Ser~Nam Lim}.} \bibinfo{year}{2021}\natexlab{}.
\newblock \showarticletitle{Large scale learning on non-homophilous graphs: New
  benchmarks and strong simple methods}.
\newblock \bibinfo{journal}{\emph{Advances in Neural Information Processing
  Systems}}  \bibinfo{volume}{34} (\bibinfo{year}{2021}),
  \bibinfo{pages}{20887--20902}.
\newblock


\bibitem[Lin et~al\mbox{.}(2015)]%
        {lin2015learning}
\bibfield{author}{\bibinfo{person}{Yankai Lin}, \bibinfo{person}{Zhiyuan Liu},
  \bibinfo{person}{Maosong Sun}, \bibinfo{person}{Yang Liu}, {and}
  \bibinfo{person}{Xuan Zhu}.} \bibinfo{year}{2015}\natexlab{}.
\newblock \showarticletitle{Learning entity and relation embeddings for
  knowledge graph completion}. In \bibinfo{booktitle}{\emph{Proceedings of the
  AAAI conference on artificial intelligence}}, Vol.~\bibinfo{volume}{29}.
\newblock


\bibitem[Liu et~al\mbox{.}(2019)]%
        {liu2019log2vec}
\bibfield{author}{\bibinfo{person}{Fucheng Liu}, \bibinfo{person}{Yu Wen},
  \bibinfo{person}{Dongxue Zhang}, \bibinfo{person}{Xihe Jiang},
  \bibinfo{person}{Xinyu Xing}, {and} \bibinfo{person}{Dan Meng}.}
  \bibinfo{year}{2019}\natexlab{}.
\newblock \showarticletitle{Log2vec: A heterogeneous graph embedding based
  approach for detecting cyber threats within enterprise}. In
  \bibinfo{booktitle}{\emph{Proceedings of the 2019 ACM SIGSAC conference on
  computer and communications security}}. \bibinfo{pages}{1777--1794}.
\newblock


\bibitem[Liu et~al\mbox{.}(2018)]%
        {liu2018towards}
\bibfield{author}{\bibinfo{person}{Yushan Liu}, \bibinfo{person}{Mu Zhang},
  \bibinfo{person}{Ding Li}, \bibinfo{person}{Kangkook Jee},
  \bibinfo{person}{Zhichun Li}, \bibinfo{person}{Zhenyu Wu},
  \bibinfo{person}{Junghwan Rhee}, {and} \bibinfo{person}{Prateek Mittal}.}
  \bibinfo{year}{2018}\natexlab{}.
\newblock \showarticletitle{Towards a Timely Causality Analysis for Enterprise
  Security.}. In \bibinfo{booktitle}{\emph{NDSS}}.
\newblock


\bibitem[M{\k{a}}dry et~al\mbox{.}(2017)]%
        {mkadry2017towards}
\bibfield{author}{\bibinfo{person}{Aleksander M{\k{a}}dry},
  \bibinfo{person}{Aleksandar Makelov}, \bibinfo{person}{Ludwig Schmidt},
  \bibinfo{person}{Dimitris Tsipras}, {and} \bibinfo{person}{Adrian Vladu}.}
  \bibinfo{year}{2017}\natexlab{}.
\newblock \showarticletitle{Towards deep learning models resistant to
  adversarial attacks}.
\newblock \bibinfo{journal}{\emph{stat}} \bibinfo{volume}{1050},
  \bibinfo{number}{9} (\bibinfo{year}{2017}).
\newblock


\bibitem[Manzoor et~al\mbox{.}(2016)]%
        {manzoor2016fast}
\bibfield{author}{\bibinfo{person}{Emaad Manzoor}, \bibinfo{person}{Sadegh~M
  Milajerdi}, {and} \bibinfo{person}{Leman Akoglu}.}
  \bibinfo{year}{2016}\natexlab{}.
\newblock \showarticletitle{Fast memory-efficient anomaly detection in
  streaming heterogeneous graphs}. In \bibinfo{booktitle}{\emph{Proceedings of
  the 22nd ACM SIGKDD international conference on knowledge discovery and data
  mining}}. \bibinfo{pages}{1035--1044}.
\newblock


\bibitem[Mikolov(2013)]%
        {mikolov2013efficient}
\bibfield{author}{\bibinfo{person}{Tomas Mikolov}.}
  \bibinfo{year}{2013}\natexlab{}.
\newblock \showarticletitle{Efficient estimation of word representations in
  vector space}.
\newblock \bibinfo{journal}{\emph{arXiv preprint arXiv:1301.3781}}
  (\bibinfo{year}{2013}).
\newblock


\bibitem[Milajerdi et~al\mbox{.}(2019a)]%
        {milajerdi2019poirot}
\bibfield{author}{\bibinfo{person}{Sadegh~M Milajerdi},
  \bibinfo{person}{Birhanu Eshete}, \bibinfo{person}{Rigel Gjomemo}, {and}
  \bibinfo{person}{VN Venkatakrishnan}.} \bibinfo{year}{2019}\natexlab{a}.
\newblock \showarticletitle{Poirot: Aligning attack behavior with kernel audit
  records for cyber threat hunting}. In \bibinfo{booktitle}{\emph{Proceedings
  of the 2019 ACM SIGSAC conference on computer and communications security}}.
  \bibinfo{pages}{1795--1812}.
\newblock


\bibitem[Milajerdi et~al\mbox{.}(2019b)]%
        {milajerdi2019holmes}
\bibfield{author}{\bibinfo{person}{Sadegh~M Milajerdi}, \bibinfo{person}{Rigel
  Gjomemo}, \bibinfo{person}{Birhanu Eshete}, \bibinfo{person}{Ramachandran
  Sekar}, {and} \bibinfo{person}{VN Venkatakrishnan}.}
  \bibinfo{year}{2019}\natexlab{b}.
\newblock \showarticletitle{Holmes: real-time apt detection through correlation
  of suspicious information flows}. In \bibinfo{booktitle}{\emph{2019 IEEE
  Symposium on Security and Privacy (SP)}}. IEEE, \bibinfo{pages}{1137--1152}.
\newblock


\bibitem[Ofori-Boateng et~al\mbox{.}(2021)]%
        {ofori2021topological}
\bibfield{author}{\bibinfo{person}{Dorcas Ofori-Boateng},
  \bibinfo{person}{I~Segovia Dominguez}, \bibinfo{person}{C Akcora},
  \bibinfo{person}{Murat Kantarcioglu}, {and} \bibinfo{person}{Yulia~R Gel}.}
  \bibinfo{year}{2021}\natexlab{}.
\newblock \showarticletitle{Topological anomaly detection in dynamic multilayer
  blockchain networks}. In \bibinfo{booktitle}{\emph{Machine Learning and
  Knowledge Discovery in Databases. Research Track: European Conference, ECML
  PKDD 2021, Bilbao, Spain, September 13--17, 2021, Proceedings, Part I 21}}.
  Springer, \bibinfo{pages}{788--804}.
\newblock


\bibitem[Paccagnella et~al\mbox{.}(2020)]%
        {paccagnella2020custos}
\bibfield{author}{\bibinfo{person}{Riccardo Paccagnella},
  \bibinfo{person}{Pubali Datta}, \bibinfo{person}{Wajih~Ul Hassan},
  \bibinfo{person}{Adam Bates}, \bibinfo{person}{Christopher Fletcher},
  \bibinfo{person}{Andrew Miller}, {and} \bibinfo{person}{Dave Tian}.}
  \bibinfo{year}{2020}\natexlab{}.
\newblock \showarticletitle{Custos: Practical tamper-evident auditing of
  operating systems using trusted execution}. In
  \bibinfo{booktitle}{\emph{Network and distributed system security
  symposium}}.
\newblock


\bibitem[Pasquier et~al\mbox{.}(2017)]%
        {pasquier2017practical}
\bibfield{author}{\bibinfo{person}{Thomas Pasquier}, \bibinfo{person}{Xueyuan
  Han}, \bibinfo{person}{Mark Goldstein}, \bibinfo{person}{Thomas Moyer},
  \bibinfo{person}{David Eyers}, \bibinfo{person}{Margo Seltzer}, {and}
  \bibinfo{person}{Jean Bacon}.} \bibinfo{year}{2017}\natexlab{}.
\newblock \showarticletitle{Practical whole-system provenance capture}. In
  \bibinfo{booktitle}{\emph{Proceedings of the 2017 Symposium on Cloud
  Computing}}. \bibinfo{pages}{405--418}.
\newblock


\bibitem[Pasquier et~al\mbox{.}(2018)]%
        {pasquier2018runtime}
\bibfield{author}{\bibinfo{person}{Thomas Pasquier}, \bibinfo{person}{Xueyuan
  Han}, \bibinfo{person}{Thomas Moyer}, \bibinfo{person}{Adam Bates},
  \bibinfo{person}{Olivier Hermant}, \bibinfo{person}{David Eyers},
  \bibinfo{person}{Jean Bacon}, {and} \bibinfo{person}{Margo Seltzer}.}
  \bibinfo{year}{2018}\natexlab{}.
\newblock \showarticletitle{Runtime analysis of whole-system provenance}. In
  \bibinfo{booktitle}{\emph{Proceedings of the 2018 ACM SIGSAC conference on
  computer and communications security}}. \bibinfo{pages}{1601--1616}.
\newblock


\bibitem[Pei et~al\mbox{.}(2016)]%
        {pei2016hercule}
\bibfield{author}{\bibinfo{person}{Kexin Pei}, \bibinfo{person}{Zhongshu Gu},
  \bibinfo{person}{Brendan Saltaformaggio}, \bibinfo{person}{Shiqing Ma},
  \bibinfo{person}{Fei Wang}, \bibinfo{person}{Zhiwei Zhang},
  \bibinfo{person}{Luo Si}, \bibinfo{person}{Xiangyu Zhang}, {and}
  \bibinfo{person}{Dongyan Xu}.} \bibinfo{year}{2016}\natexlab{}.
\newblock \showarticletitle{Hercule: Attack story reconstruction via community
  discovery on correlated log graph}. In \bibinfo{booktitle}{\emph{Proceedings
  of the 32Nd Annual Conference on Computer Security Applications}}.
  \bibinfo{pages}{583--595}.
\newblock


\bibitem[Peng et~al\mbox{.}(2021)]%
        {peng2021reinforced}
\bibfield{author}{\bibinfo{person}{Hao Peng}, \bibinfo{person}{Ruitong Zhang},
  \bibinfo{person}{Yingtong Dou}, \bibinfo{person}{Renyu Yang},
  \bibinfo{person}{Jingyi Zhang}, {and} \bibinfo{person}{Philip~S Yu}.}
  \bibinfo{year}{2021}\natexlab{}.
\newblock \showarticletitle{Reinforced neighborhood selection guided
  multi-relational graph neural networks}.
\newblock \bibinfo{journal}{\emph{ACM Transactions on Information Systems
  (TOIS)}} \bibinfo{volume}{40}, \bibinfo{number}{4} (\bibinfo{year}{2021}),
  \bibinfo{pages}{1--46}.
\newblock


\bibitem[Popescu et~al\mbox{.}(2009)]%
        {popescu2009multilayer}
\bibfield{author}{\bibinfo{person}{Marius-Constantin Popescu},
  \bibinfo{person}{Valentina~E Balas}, \bibinfo{person}{Liliana
  Perescu-Popescu}, {and} \bibinfo{person}{Nikos Mastorakis}.}
  \bibinfo{year}{2009}\natexlab{}.
\newblock \showarticletitle{Multilayer perceptron and neural networks}.
\newblock \bibinfo{journal}{\emph{WSEAS Transactions on Circuits and Systems}}
  \bibinfo{volume}{8}, \bibinfo{number}{7} (\bibinfo{year}{2009}),
  \bibinfo{pages}{579--588}.
\newblock


\bibitem[Rehman et~al\mbox{.}(2024)]%
        {rehman2024flash}
\bibfield{author}{\bibinfo{person}{Mati~Ur Rehman}, \bibinfo{person}{Hadi
  Ahmadi}, {and} \bibinfo{person}{Wajih~Ul Hassan}.}
  \bibinfo{year}{2024}\natexlab{}.
\newblock \showarticletitle{FLASH: A Comprehensive Approach to Intrusion
  Detection via Provenance Graph Representation Learning}. In
  \bibinfo{booktitle}{\emph{2024 IEEE Symposium on Security and Privacy (SP)}}.
  IEEE Computer Society, \bibinfo{pages}{139--139}.
\newblock


\bibitem[{SektorCERT}(2023)]%
        {sektorcert2023attack}
\bibfield{author}{\bibinfo{person}{{SektorCERT}}.}
  \bibinfo{year}{2023}\natexlab{}.
\newblock \bibinfo{title}{The attack against Danish critical infrastructure}.
\newblock \bibinfo{howpublished}{Available online}.
\newblock
\urldef\tempurl%
\url{https://sektorcert.dk/wp-content/uploads/2023/11/SektorCERT-The-attack-against-Danish-critical-infrastructure-TLP-CLEAR.pdf}
\showURL{%
\tempurl}


\bibitem[Sharma et~al\mbox{.}(2023)]%
        {sharma2023advanced}
\bibfield{author}{\bibinfo{person}{Amit Sharma}, \bibinfo{person}{Brij~B
  Gupta}, \bibinfo{person}{Awadhesh~Kumar Singh}, {and} \bibinfo{person}{VK
  Saraswat}.} \bibinfo{year}{2023}\natexlab{}.
\newblock \showarticletitle{Advanced persistent threats (apt): evolution,
  anatomy, attribution and countermeasures}.
\newblock \bibinfo{journal}{\emph{Journal of Ambient Intelligence and Humanized
  Computing}} \bibinfo{volume}{14}, \bibinfo{number}{7} (\bibinfo{year}{2023}),
  \bibinfo{pages}{9355--9381}.
\newblock


\bibitem[Sun et~al\mbox{.}(2021)]%
        {sun2021sugar}
\bibfield{author}{\bibinfo{person}{Qingyun Sun}, \bibinfo{person}{Jianxin Li},
  \bibinfo{person}{Hao Peng}, \bibinfo{person}{Jia Wu},
  \bibinfo{person}{Yuanxing Ning}, \bibinfo{person}{Philip~S Yu}, {and}
  \bibinfo{person}{Lifang He}.} \bibinfo{year}{2021}\natexlab{}.
\newblock \showarticletitle{Sugar: Subgraph neural network with reinforcement
  pooling and self-supervised mutual information mechanism}. In
  \bibinfo{booktitle}{\emph{Proceedings of the web conference 2021}}.
  \bibinfo{pages}{2081--2091}.
\newblock


\bibitem[Team(2023)]%
        {symantec2023grayling}
\bibfield{author}{\bibinfo{person}{Threat~Hunter Team}.}
  \bibinfo{year}{2023}\natexlab{}.
\newblock \bibinfo{booktitle}{\emph{Grayling: Previously Unseen Threat Actor
  Targets Multiple Organizations in Taiwan}}.
\newblock
\urldef\tempurl%
\url{https://symantec-enterprise-blogs.security.com/threat-intelligence/grayling-taiwan-cyber-attacks}
\showURL{%
\tempurl}


\bibitem[Tian et~al\mbox{.}(2024)]%
        {tian2024survey}
\bibfield{author}{\bibinfo{person}{Songsong Tian}, \bibinfo{person}{Lusi Li},
  \bibinfo{person}{Weijun Li}, \bibinfo{person}{Hang Ran}, \bibinfo{person}{Xin
  Ning}, {and} \bibinfo{person}{Prayag Tiwari}.}
  \bibinfo{year}{2024}\natexlab{}.
\newblock \showarticletitle{A survey on few-shot class-incremental learning}.
\newblock \bibinfo{journal}{\emph{Neural Networks}}  \bibinfo{volume}{169}
  (\bibinfo{year}{2024}), \bibinfo{pages}{307--324}.
\newblock


\bibitem[Tramer and Boneh(2019)]%
        {tramer2019adversarial}
\bibfield{author}{\bibinfo{person}{Florian Tramer} {and} \bibinfo{person}{Dan
  Boneh}.} \bibinfo{year}{2019}\natexlab{}.
\newblock \showarticletitle{Adversarial training and robustness for multiple
  perturbations}.
\newblock \bibinfo{journal}{\emph{Advances in neural information processing
  systems}}  \bibinfo{volume}{32} (\bibinfo{year}{2019}).
\newblock


\bibitem[Uc-Cetina et~al\mbox{.}(2023)]%
        {uc2023survey}
\bibfield{author}{\bibinfo{person}{Victor Uc-Cetina},
  \bibinfo{person}{Nicol{\'a}s Navarro-Guerrero}, \bibinfo{person}{Anabel
  Martin-Gonzalez}, \bibinfo{person}{Cornelius Weber}, {and}
  \bibinfo{person}{Stefan Wermter}.} \bibinfo{year}{2023}\natexlab{}.
\newblock \showarticletitle{Survey on reinforcement learning for language
  processing}.
\newblock \bibinfo{journal}{\emph{Artificial Intelligence Review}}
  \bibinfo{volume}{56}, \bibinfo{number}{2} (\bibinfo{year}{2023}),
  \bibinfo{pages}{1543--1575}.
\newblock


\bibitem[Veli{\v{c}}kovi{\'c} et~al\mbox{.}(2017)]%
        {velivckovic2017graph}
\bibfield{author}{\bibinfo{person}{Petar Veli{\v{c}}kovi{\'c}},
  \bibinfo{person}{Guillem Cucurull}, \bibinfo{person}{Arantxa Casanova},
  \bibinfo{person}{Adriana Romero}, \bibinfo{person}{Pietro Lio}, {and}
  \bibinfo{person}{Yoshua Bengio}.} \bibinfo{year}{2017}\natexlab{}.
\newblock \showarticletitle{Graph attention networks}.
\newblock \bibinfo{journal}{\emph{arXiv preprint arXiv:1710.10903}}
  (\bibinfo{year}{2017}).
\newblock


\bibitem[Wang et~al\mbox{.}(2020)]%
        {wang2020you}
\bibfield{author}{\bibinfo{person}{Qi Wang}, \bibinfo{person}{Wajih~Ul Hassan},
  \bibinfo{person}{Ding Li}, \bibinfo{person}{Kangkook Jee},
  \bibinfo{person}{Xiao Yu}, \bibinfo{person}{Kexuan Zou},
  \bibinfo{person}{Junghwan Rhee}, \bibinfo{person}{Zhengzhang Chen},
  \bibinfo{person}{Wei Cheng}, \bibinfo{person}{Carl~A Gunter},
  {et~al\mbox{.}}} \bibinfo{year}{2020}\natexlab{}.
\newblock \showarticletitle{You Are What You Do: Hunting Stealthy Malware via
  Data Provenance Analysis.}. In \bibinfo{booktitle}{\emph{NDSS}}.
\newblock


\bibitem[Wang et~al\mbox{.}(2022)]%
        {wang2022threatrace}
\bibfield{author}{\bibinfo{person}{Su Wang}, \bibinfo{person}{Zhiliang Wang},
  \bibinfo{person}{Tao Zhou}, \bibinfo{person}{Hongbin Sun},
  \bibinfo{person}{Xia Yin}, \bibinfo{person}{Dongqi Han}, \bibinfo{person}{Han
  Zhang}, \bibinfo{person}{Xingang Shi}, {and} \bibinfo{person}{Jiahai Yang}.}
  \bibinfo{year}{2022}\natexlab{}.
\newblock \showarticletitle{Threatrace: Detecting and tracing host-based
  threats in node level through provenance graph learning}.
\newblock \bibinfo{journal}{\emph{IEEE Transactions on Information Forensics
  and Security}}  \bibinfo{volume}{17} (\bibinfo{year}{2022}),
  \bibinfo{pages}{3972--3987}.
\newblock


\bibitem[Wen et~al\mbox{.}(2020)]%
        {wen2020asa}
\bibfield{author}{\bibinfo{person}{Rui Wen}, \bibinfo{person}{Jianyu Wang},
  \bibinfo{person}{Chunming Wu}, {and} \bibinfo{person}{Jian Xiong}.}
  \bibinfo{year}{2020}\natexlab{}.
\newblock \showarticletitle{Asa: Adversary situation awareness via
  heterogeneous graph convolutional networks}. In
  \bibinfo{booktitle}{\emph{Companion Proceedings of the Web Conference 2020}}.
  \bibinfo{pages}{674--678}.
\newblock


\bibitem[Xie et~al\mbox{.}(2022)]%
        {xie2022efficiency}
\bibfield{author}{\bibinfo{person}{Zhiwen Xie}, \bibinfo{person}{Runjie Zhu},
  \bibinfo{person}{Jin Liu}, \bibinfo{person}{Guangyou Zhou}, {and}
  \bibinfo{person}{Jimmy~Xiangji Huang}.} \bibinfo{year}{2022}\natexlab{}.
\newblock \showarticletitle{An efficiency relation-specific graph
  transformation network for knowledge graph representation learning}.
\newblock \bibinfo{journal}{\emph{Information Processing \& Management}}
  \bibinfo{volume}{59}, \bibinfo{number}{6} (\bibinfo{year}{2022}),
  \bibinfo{pages}{103076}.
\newblock


\bibitem[Yang et~al\mbox{.}(2023)]%
        {yang2023prographer}
\bibfield{author}{\bibinfo{person}{Fan Yang}, \bibinfo{person}{Jiacen Xu},
  \bibinfo{person}{Chunlin Xiong}, \bibinfo{person}{Zhou Li}, {and}
  \bibinfo{person}{Kehuan Zhang}.} \bibinfo{year}{2023}\natexlab{}.
\newblock \showarticletitle{$\{$PROGRAPHER$\}$: An Anomaly Detection System
  based on Provenance Graph Embedding}. In \bibinfo{booktitle}{\emph{32nd
  USENIX Security Symposium (USENIX Security 23)}}.
  \bibinfo{pages}{4355--4372}.
\newblock


\bibitem[Yuan et~al\mbox{.}(2021)]%
        {yuan2021explainability}
\bibfield{author}{\bibinfo{person}{Hao Yuan}, \bibinfo{person}{Haiyang Yu},
  \bibinfo{person}{Jie Wang}, \bibinfo{person}{Kang Li}, {and}
  \bibinfo{person}{Shuiwang Ji}.} \bibinfo{year}{2021}\natexlab{}.
\newblock \showarticletitle{On explainability of graph neural networks via
  subgraph explorations}. In \bibinfo{booktitle}{\emph{International conference
  on machine learning}}. PMLR, \bibinfo{pages}{12241--12252}.
\newblock


\bibitem[Zengy et~al\mbox{.}(2022)]%
        {zengy2022shadewatcher}
\bibfield{author}{\bibinfo{person}{Jun Zengy}, \bibinfo{person}{Xiang Wang},
  \bibinfo{person}{Jiahao Liu}, \bibinfo{person}{Yinfang Chen},
  \bibinfo{person}{Zhenkai Liang}, \bibinfo{person}{Tat-Seng Chua}, {and}
  \bibinfo{person}{Zheng~Leong Chua}.} \bibinfo{year}{2022}\natexlab{}.
\newblock \showarticletitle{Shadewatcher: Recommendation-guided cyber threat
  analysis using system audit records}. In \bibinfo{booktitle}{\emph{2022 IEEE
  Symposium on Security and Privacy (SP)}}. IEEE, \bibinfo{pages}{489--506}.
\newblock


\bibitem[Zhang et~al\mbox{.}(2022)]%
        {zhang2022hierarchical}
\bibfield{author}{\bibinfo{person}{Zaixi Zhang}, \bibinfo{person}{Qi Liu},
  \bibinfo{person}{Qingyong Hu}, {and} \bibinfo{person}{Chee-Kong Lee}.}
  \bibinfo{year}{2022}\natexlab{}.
\newblock \showarticletitle{Hierarchical graph transformer with adaptive node
  sampling}.
\newblock \bibinfo{journal}{\emph{Advances in Neural Information Processing
  Systems}}  \bibinfo{volume}{35} (\bibinfo{year}{2022}),
  \bibinfo{pages}{21171--21183}.
\newblock


\bibitem[Zheleva and Getoor(2009)]%
        {zheleva2009join}
\bibfield{author}{\bibinfo{person}{Elena Zheleva} {and} \bibinfo{person}{Lise
  Getoor}.} \bibinfo{year}{2009}\natexlab{}.
\newblock \showarticletitle{To join or not to join: the illusion of privacy in
  social networks with mixed public and private user profiles}. In
  \bibinfo{booktitle}{\emph{Proceedings of the 18th international conference on
  World wide web}}. \bibinfo{pages}{531--540}.
\newblock


\bibitem[Zhu et~al\mbox{.}(2020)]%
        {zhu2020beyond}
\bibfield{author}{\bibinfo{person}{Jiong Zhu}, \bibinfo{person}{Yujun Yan},
  \bibinfo{person}{Lingxiao Zhao}, \bibinfo{person}{Mark Heimann},
  \bibinfo{person}{Leman Akoglu}, {and} \bibinfo{person}{Danai Koutra}.}
  \bibinfo{year}{2020}\natexlab{}.
\newblock \showarticletitle{Beyond homophily in graph neural networks: Current
  limitations and effective designs}.
\newblock \bibinfo{journal}{\emph{Advances in neural information processing
  systems}}  \bibinfo{volume}{33} (\bibinfo{year}{2020}),
  \bibinfo{pages}{7793--7804}.
\newblock


\end{thebibliography}

\end{document}